\documentclass[journal=jctcce,manuscript=article,layout=twocolumn]{achemso}

\usepackage[colorlinks,linkcolor=blue,citecolor=blue,urlcolor=black,bookmarks=false,hypertexnames=true]{hyperref} 

\usepackage[version=3]{mhchem}
\usepackage{amssymb}
\usepackage{color}
\usepackage{braket}
\usepackage{xspace}
\usepackage{cleveref}
\usepackage{graphicx}
\usepackage{subfigure}
\usepackage{threeparttable}
\usepackage{textcomp}
\usepackage{setspace}
\usepackage{caption}
\usepackage{comment}
\usepackage{lmodern}

\usepackage{array}
\newcolumntype{L}[1]{>{\raggedright\let\newline\\\arraybackslash\hspace{0pt}}m{#1}}
\newcolumntype{C}[1]{>{\centering\let\newline\\\arraybackslash\hspace{0pt}}m{#1}}
\newcolumntype{R}[1]{>{\raggedleft\let\newline\\\arraybackslash\hspace{0pt}}m{#1}}

\makeatletter
\let\l@addto@macro\relax
\makeatother
\usepackage[fontsize=11pt]{scrextend}

\let\oldmaketitle\maketitle
\let\maketitle\relax

\newcommand{\kc}{\textbf{k}}

\crefname{figure}{Figure}{Figures}
\crefname{table}{Table}{Tables}
\crefname{equation}{Eq.}{Eqs.}
\crefname{section}{Section}{Sections}
\crefname{subsection}{Section}{Sections}

\author{Samragni Banerjee}
\affiliation{%
     Department of Chemistry and Biochemistry,
     The Ohio State University,
     Columbus, Ohio 43210, United States
}
 \author{Alexander Yu.\@ Sokolov}
 \email{sokolov.8@osu.edu}
 \affiliation{%
     Department of Chemistry and Biochemistry,
     The Ohio State University,
     Columbus, Ohio 43210, United States
 }

\begin{tocentry}
\includegraphics{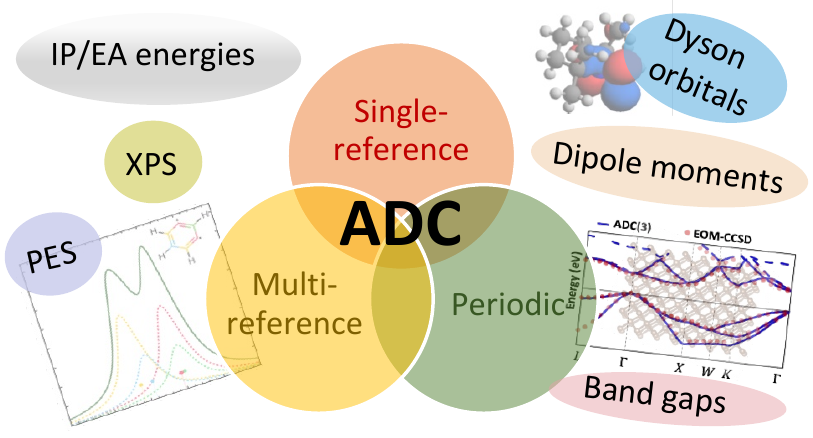}
\end{tocentry}

\title{{\color{blue}Algebraic Diagrammatic Construction Theory\\for Simulating Charged Excited States\\and Photoelectron Spectra}}

\begin{document}

\newcommand*{\abstractext}{
Charged excitations are electronic transitions that involve a change in the total charge of a molecule or material. 
Understanding the properties and reactivity of charged species requires insights from theoretical calculations that can accurately describe orbital relaxation and electron correlation effects in open-shell electronic states.
In this perspective, we review the current state of algebraic diagrammatic construction (ADC) theory for simulating charged excitations and its recent developments. 
We start with a short overview of ADC formalism for the one-particle Green's function, including its single- and multireference formulations and extension to periodic systems.
Next, we focus on the capabilities of ADC methods and discuss recent findings about their accuracy for calculating a wide range of excited-state properties. 
We conclude our perspective by outlining possible directions for future developments of this theoretical approach.
\vspace{0.25cm}
}

\twocolumn[
\begin{@twocolumnfalse}
\oldmaketitle
\vspace{-0.75cm}
\begin{abstract}
\abstractext
\end{abstract}
\end{@twocolumnfalse}
]

\section{Introduction}

Electronic excitations that change the charge state of molecules or materials are central to many important processes in chemistry, biochemistry, and materials science. \cite{Ehring:1992p472,Schermann:2007,Wannier:1937p191,Illenberger:1992p1589,Arnold:1976p5931,Bergmann:1969p158,Scuseria:2021pe2113648118,Roca:2008p095104,Li:2002p1596,Wesolowski:2001p4023,Richardson:2002p10163,Dutta:2015p753,Aflatooni:1998p6205} 
Playing a crucial role in redox chemistry and catalysis, charged excitations are also common in atmospheric and combustion chemical reactions, and are responsible for structural damage in some biomolecules.\cite{Umstead:1966p293,Uddin:2020p1905739,Steenken:1989p503,Yan:1992p1983,Colson:1995p3867,Wintjens:2000p393,Huels:1998p1309}
Additionally, charged excitations are the primary transitions probed by photoelectron spectroscopy, which reveals important details about the electronic structure of molecules and materials by measuring the binding energies of core and valence electrons.\cite{Hufner:2013,Chastain:1992p221}
Charged electronic states can be accessed via chemical oxidation or reduction, photoexcitation, or direct electron attachment or ionization in the gas phase or electrochemical cell.

Theoretical simulations are essential for investigating the electronic structure and properties of charged excited states, which are often elusive and difficult to study experimentally.
However, simulating charged excitations presents many challenges that put stringent requirements on what kind of theoretical methods can deliver accurate energies and properties of charged excited states.
Changing the charge state of a molecule or material significantly perturbs their electronic density, altering the size of orbitals and the distribution of charge carriers. 
To accurately model these effects, it is necessary to use advanced theoretical techniques that take into account orbital relaxation and electron correlation in open-shell states with complex electronic structures.

Theoretical methods for simulating charged excitations can be divided into two broad categories: 
i) the energy difference (or $\Delta$) approaches
and ii) the direct (or response) theories.
The $\Delta$ methods\cite{Bagus:1965pA619,Deutsch:1976p588,Naves:1991p2965,Schmitt:1992p1,Besley:2009p124308,Ambroise:2018p325,Aagren:1993p45,Triguero:1999p195,Duflot:2010p27,Shim:2011p5703,Ljubic:2014p2333,Su:2016p2285,South:2016p21010,Smiga:2018p4780,Holme:2011p4104,Klues:2014p014302,Zheng:2019p4945} compute charged excitation energies and properties by performing separate calculations for each electronic state, including the neutral ground state.
These simulations typically begin with a self-consistent-field ($\Delta$SCF) \cite{Jones:1989p689,Roothaan:1951p69,Bagus:1965pA619,Deutsch:1976p588,Naves:1991p2965,Schmitt:1992p1,Besley:2009p124308,Ambroise:2018p325} or complete-active-space SCF ($\Delta$CASSCF)\cite{Naves:1991p2965,Aagren:1993p45} optimization of electronic wavefunction, followed by a post-$\Delta$SCF or post-$\Delta$CASSCF treatment of dynamic electron correlation. \cite{Szabo:2012,Helgaker:2013} 
Although the $\Delta$ approaches can accurately capture relaxation of orbitals and wavefunctions upon charged excitation, their computations are only straightforward for the lowest-energy excited states of particular symmetry. 
For the higher-energy excited states, the $\Delta$SCF or $\Delta$CASSCF wavefunction optimizations are prone to convergence problems or variational collapse to lower-energy states. In addition, the wavefunctions calculated using the $\Delta$ methods may not be orthogonal to each other and their calculations become prohibitively expensive for systems with high density of states.

In contrast to the $\Delta$ methods, response theories perform a single calculation for all electronic states simultaneously, ensuring that their wavefunctions are orthogonal. These methods optimize the ground-state wavefunction and compute the excitation energies and transition properties by assuming that the electron correlation in all states (ground and excited) is similar.
Although response theories are less powerful than the $\Delta$ methods in capturing the wavefunction relaxation effects, their calculations are more efficient, do not suffer from systematic convergence problems, and can yield accurate results provided that they employ a sufficiently high level of electron correlation treatment. 
For these reasons, a wide range of response methods have been developed, including time-dependent density functional theory,\cite{Runge:1984p997,Besley:2010p12024,Akama:2010p054104,Lopata:2012p3284,Her23} propagator (or Green's function) approaches,\cite{Mckechnie:2015p194114,Nguyen:2012p081101,Hedin:1965p796,Faleev:2004p126406,vanSchilfgaarde:2006p226402,vanSetten:2013p232,Reining:2017pe1344,Linderberg:2004,Cederbaum:2007p205,Ortiz:2013p123} traditional and symmetry-adapted-cluster configuration interaction,\cite{Nakatsuji:1978p2053,Nakatsuji:1979p329,Asmuruf:2008p267,Maganas:2014p6374,Toffoli:2016p4996,Ehlert:2017p116,Kuramoto:2005p014304,Ehara:2009p195} and coupled cluster theory (CC) in linear-response\cite{Sekino:1984p255,Sauer:2009p555,Rocha:2009p44,Koch:1990p3345,Piecuch:1996p4699} or equation-of-motion\cite{Nooijen:1992p55,Nooijen:1993p15,Nooijen:1995p1681,Stanton:1993p7029,Krylov:2008p433,Kowalski:2014p094102,BhaskaranNair:2016p144101,Peng:2018p4335} formulations.

A special class of response methods is algebraic diagrammatic construction theory (ADC).\cite{Schirmer:1982p2395,Schirmer:1983p1237,Schirmer:1991p4647,Mertins:1996p2140,Schirmer:2004p11449,Dreuw:2014p82,Sokolov:2018p204113,Banerjee:2022p5337} 
Since its original formulation in 1982 by Jochen Schirmer, ADC has become one of the standard electronic structure theories for simulating neutral excited states and UV/Vis spectra of molecules, owing to the moderate computational cost and relatively high accuracy of its methods. 
Although traditionally formulated within the propagator formalism, conventional (single-reference) ADC theory has a close connection with wavefunction-based M\o ller--Plesset perturbation theory\cite{Moller:1934kp618} and equation-of-motion CC (EOM-CC),
\cite{Nooijen:1992p55,Nooijen:1993p15,Nooijen:1995p1681,Stanton:1993p7029,Krylov:2008p433,Kowalski:2014p094102,BhaskaranNair:2016p144101,Peng:2018p4335}  
offering a hierarchy of size-consistent methods that systematically improve accuracy with increasing order of approximation.
Taking advantage of noniterative and Hermitian equations, low-order ADC methods are more computationally efficient than EOM-CC with single and double excitations (EOM-CCSD) and are capable of delivering results with similar accuracy.
In addition to neutral excited states, the ADC framework has been extended to simulating charged and two-photon electronic transitions with excitation energies spanning from the visible to X-ray regions of electromagnetic spectrum. \cite{Schirmer:1983p1237,Angonoa:1987p6789,Schirmer:1998p4734,Schirmer:2001p10621,Thiel:2003p2088,Trofimov:2005p144115,Knippenberg:2012p064107,Schneider:2015p144103,Banerjee:2019p224112,Dempwolff:2019p064108,Dempwolff:2020p024113,Dempwolff:2020p024125,Dempwolff:2021p104117,Banerjee:2021p074105,Dempwolff:2022p054114,Liu:2020p174109,Stahl:2022p044106,Chatterjee:2019p5908,Chatterjee:2020p6343,Moura:2022p4769,Moura:2022p8041}

The original formulation of ADC methods for charged excitations proposed in 1983 was based on the Dyson ADC framework,\cite{Schirmer:1983p1237} which allows to simulate the electron-attached ($N+1$) and ionized ($N-1$) states in one calculation by approximately solving the Dyson equation. 
In 1998, Schirmer, Trofimov, and Stelter formulated the non-Dyson ADC formalism \cite{Schirmer:1998p4734} that enables calculating the $N+1$ or $N-1$ excited states independently from each other, bypassing the solution of Dyson equation in a manner similar to Green's function methods based on coupled cluster theory\cite{Prasad:1985p1287,Mukherjee:1989p257,Nooijen:1992p55,Nooijen:1993p15,Nooijen:1995p1681,Meissner:1993p67} developed earlier.
Although the non-Dyson ADC methods are more computationally efficient than those based on the Dyson ADC framework, the number of studies reporting their calculations remained small until 2019 when these methods received a renewed interest.\cite{Banerjee:2019p224112,Dempwolff:2019p064108,Dempwolff:2020p024113,Dempwolff:2020p024125,Dempwolff:2021p104117,Banerjee:2021p074105,Dempwolff:2022p054114}

In this perspective, we review the current state of non-Dyson ADC theory for charged excitations, including the recent progress in single-reference ADC for molecules and periodic materials, and the development of multireference ADC for molecular systems with challenging electronic structure.
We begin with a brief overview of theoretical background behind ADC and its approximations (\cref{sec:theoretical_background}).
We then discuss the capabilities of ADC methods and summarize recent findings about their accuracy (\cref{sec:capabilities_accuracy}).
Finally, we conclude our perspective by discussing possible directions for future developments (\cref{sec:summary_outlook}). 

\section{Theoretical Background}
\label{sec:theoretical_background}

\subsection{One-particle Green's function}
\label{sec:theoretical_background:1_gf}

The central mathematical object in the ADC theory of charged excitations is the one-particle Green's function (or so-called single-particle propagator, 1-GF), which for an $N$-electron chemical system contains information about the energies and properties of all electron-attached ($N+1$) and ionized ($N-1$) states.\cite{Fetter:1971quantum,Dickhoff:2008many,Schirmer:2018} 
1-GF can be expressed as an expectation value 
\begin{align}
	\label{eq:time_GF}
	G_{pq}(t,t') = -i\braket{ \Psi_{0}^{N} | \mathcal{T} [ a_{p}(t) a_{q}^{\dagger}(t') ] | \Psi_{0}^{N}}
\end{align}
where $\ket{\Psi_{0}^{N}}$ is the $N$-electron ground-state wavefunction, $a_{q}^{\dagger}(t')$ is a creation operator that adds an electron to spin-orbital $\ket{\psi_q}$ at time $t'$, $a_{p}(t)$ is an annihilation operator that removes an electron from spin-orbital $\ket{\psi_p}$ at a different time $t$, and $\mathcal{T}$ is the operator responsible for time-ordering.
\cref{eq:time_GF} describes both the forward ($t > t'$) and backward ($t < t'$) processes in time corresponding to the one-electron attachment and ionization of the ground electronic state, respectively.

By carrying out a Fourier transformation of \cref{eq:time_GF}, the 1-GF can be also defined in the frequency domain as 
\begin{align}
\label{eq:freq_GF}
G_{pq}(\omega) 
               &=\langle\Psi_{0}^{N}|a_{p} (\omega - H + E_{0}^{N})^{-1}a_{q}^{\dagger}|\Psi_{0}^{N}\rangle\notag\\
               &+ \langle\Psi_{0}^{N}|a_{q}^{\dagger} (\omega + H - E_{0}^{N})^{-1} a_{p}|\Psi_{0}^{N}\rangle
\end{align}
where $H$ is the electronic Hamiltonian, $E_{0}^{N}$ is the ground-state energy, and $\omega$ is the frequency of incident radiation.
Alternatively, the frequency-dependent 1-GF can be expressed in the spectral (or Lehmann) representation \cite{Kobe:1962p448}
\begin{align}
\label{eq:freq_GF_spectral}
	G_{pq}(\omega) &= \sum_{n}\frac{\langle\Psi_{0}^{N}|a_{p}|\Psi_{n}^{N+1}\rangle \langle\Psi_{n}^{N+1}|a_{q}^{\dagger}|\Psi_{0}^{N}\rangle}{\omega - E_{n}^{N+1}+ E_{0}^{N}}\notag\\
	&+  \sum_{n}\frac{\langle\Psi_{0}^{N}|a_{q}^{\dagger}|\Psi_{n}^{N-1}\rangle \langle\Psi_{n}^{N-1}|a_{p}|\Psi_{0}^{N}\rangle}{\omega + E_{n}^{N-1} - E_{0}^{N}} \notag\\
    &\equiv G_{pq}^{+}(\omega) + G_{pq}^{-}(\omega)
\end{align}
where $G_{pq}^{+}(\omega)$ and $G_{pq}^{-}(\omega)$ are the forward and backward components of 1-GF describing electron attachment and detachment processes, respectively, $\ket{\Psi_{n}^{N+1}}$ and $\ket{\Psi_{n}^{N-1}}$ are the exact eigenstates of $(N+1)$- and $(N-1)$-electron system with energies $E_{n}^{N+1}$ and $E_{n}^{N-1}$.

\cref{eq:freq_GF_spectral} shows that 1-GF contains information about the vertical electron attachment ($E_{n}^{N+1} - E_{0}^{N}$) and ionization ($E_{0}^{N} - E_{n}^{N - 1}$) energies, as well as the corresponding transition probabilities ($|\langle\Psi_{0}^{N}|a_{p}|\Psi_{n}^{N+1}\rangle|^2$ and $|\langle\Psi_{0}^{N}|a_{q}^{\dagger}|\Psi_{n}^{N-1}\rangle|^2$).
For each component of 1-GF, the Lehmann representation can be compactly written in a matrix form as:
\begin{align}
\label{eq:freq_GF_spectral_matrix}
	\textbf{G}_{\pm}(\omega) = \mathbf{\tilde{X}}_{\pm}(\omega\textbf{1} - \boldsymbol{\tilde{\Omega}}_{\pm})^{-1}\mathbf{\tilde{X}}_{\pm}^{\dagger}
\end{align}
where $\boldsymbol{\tilde{\Omega}}_{\pm}$ are the diagonal matrices of exact vertical attachment ($\tilde{\Omega}_{+n} = E_{n}^{N+1} - E_{0}^{N}$) and ionization ($\tilde{\Omega}_{-n} = E_{0}^{N} - E_{n}^{N - 1}$) energies, while $\mathbf{\tilde{X}}_{\pm}$ are the matrices of spectroscopic amplitudes with elements $\tilde{X}_{+pn} = \langle\Psi_{0}^{N}|a_{p}|\Psi_{n}^{N+1}\rangle$ and $\tilde{X}_{-qn} = \langle\Psi_{0}^{N}|a_{q}^{\dagger}|\Psi_{n}^{N-1}\rangle$.

\subsection{General overview of ADC for charged excitations}
\label{sec:theoretical_background:gen_overview}

Although informative, \cref{eq:freq_GF_spectral} is not very useful for simulating charged excited states, since it assumes that the eigenstates $\ket{\Psi_{n}^{N+1}}$ and $\ket{\Psi_{n}^{N-1}}$ and their energies $E_{n}^{N+1}$ and $E_{n}^{N-1}$ are already known exactly! 
Calculating {\it approximate} charged excitation energies and transition probabilities by making approximations to 1-GF is the domain of propagator theory.
Algebraic diagrammatic construction (ADC) is a particular propagator approach that obtains closed-form algebraic expressions for the 1-GF (or, in general, any propagator) approximated using low-order perturbation theory. \cite{Schirmer:1983p1237,Schirmer:1998p4734,Angonoa:1987p6789,Schirmer:2001p10621,Thiel:2003p2088,Trofimov:2005p144115,Knippenberg:2012p064107,Schneider:2015p144103,Banerjee:2019p224112,Dempwolff:2019p064108,Dempwolff:2020p024113,Dempwolff:2020p024125,Dempwolff:2021p104117,Banerjee:2021p074105,Dempwolff:2022p054114,Liu:2020p174109,Stahl:2022p044106,Chatterjee:2019p5908,Chatterjee:2020p6343,Moura:2022p4769,Moura:2022p8041}

Two ADC variants for simulating charged excitations have been proposed.
In the first one called the Dyson ADC framework, \cite{Schirmer:1983p1237,Angonoa:1987p6789,Santra:2009p013002} the exact 1-GF in \cref{eq:freq_GF_spectral} is expressed in terms of the Green's function of a noninteracting system ($\mathbf{G^{(0)}}(\omega)$) via the Dyson equation\cite{Fetter:1971quantum,Dickhoff:2008many,Schirmer:2018}
\begin{align}
\label{eq:Dyson}
	\mathbf{G}(\omega) 
	&= \mathbf{G^{(0)}}(\omega) + \mathbf{G^{(0)}}(\omega)\mathbf{\Sigma}(\omega)\mathbf{G}(\omega) \notag \\
	&= \mathbf{G^{(0)}}(\omega) 
	+ \mathbf{G^{(0)}}(\omega)\mathbf{\Sigma}(\omega)\mathbf{G^{(0)}}(\omega) \notag \\
	&+ \mathbf{G^{(0)}}(\omega)\mathbf{\Sigma}(\omega)\mathbf{G^{(0)}}(\omega)\mathbf{\Sigma}(\omega)\mathbf{G^{(0)}}(\omega)
	+\ldots
\end{align}
where $\mathbf{\Sigma}(\omega)$ is a frequency-dependent potential called self-energy, which contains information about all electron correlation effects in $\mathbf{G}(\omega)$ that are not included in $\mathbf{G^{(0)}}(\omega)$.
In Dyson ADC, $\mathbf{\Sigma}(\omega)$ is expressed as a sum of static ($\omega$-independent, $\mathbf{\Sigma_s}(\infty)$) and dynamic ($\omega$-dependent, $\mathbf{\Sigma_d}(\omega)$) contributions:
\begin{align}
	\label{eq:full_self_energy}
	\mathbf{\Sigma}(\omega) = \mathbf{\Sigma_s}(\infty) + \mathbf{\Sigma_d}(\omega)
\end{align}
The dynamic self-energy $\mathbf{\Sigma_d}(\omega)$ is expanded in a perturbative series, which is truncated at a low order $n$:
\begin{align}
	\label{eq:self_energy_approx}
	\mathbf{\Sigma_d}(\omega) \approx \mathbf{\Sigma_d^{(0)}}(\omega) + \mathbf{\Sigma_d^{(1)}}(\omega) + \ldots + \mathbf{\Sigma_d^{(n)}}(\omega)
\end{align}
Taking advantage of the non-interacting nature of $\mathbf{G^{(0)}}(\omega)$ (i.e., $\mathbf{G^{(0)}}(\omega)$ corresponding to a Slater determinant wavefunction), Dyson ADC provides algebraic expressions for calculating the contributions to $\mathbf{\Sigma_d}(\omega)$ at each order in perturbation theory. 
The resulting approximate $\mathbf{\Sigma_d}(\omega)$ is used to obtain the static self-energy $\mathbf{\Sigma_s}(\infty)$ in \cref{eq:full_self_energy}, solve the Dyson equation \eqref{eq:Dyson} for approximate $\mathbf{G}(\omega)$, and compute the corresponding charged excitation energies and transition probabilities.
Since $\mathbf{G}(\omega)$ in \cref{eq:Dyson} is a sum of forward and backward components ($\textbf{G}(\omega) = \textbf{G}_{+}(\omega) + \textbf{G}_{-}(\omega)$), in Dyson ADC the electron-attached and ionized states are calculated simultaneously, in a coupled basis of $(N+1)$- and $(N-1)$-electron configurations.
In this respect, Dyson ADC is different from wavefunction-based theories (such as EOM-CC or $\Delta$ methods), which simulate the electron-attached and ionized states independently from each other.

In the second approach, termed non-Dyson ADC, \cite{Schirmer:1998p4734,Schirmer:2001p10621,Thiel:2003p2088,Trofimov:2005p144115,Schneider:2015p144103,Banerjee:2019p224112,Dempwolff:2019p064108,Dempwolff:2020p024113,Dempwolff:2020p024125,Dempwolff:2021p104117,Banerjee:2021p074105,Chatterjee:2019p5908,Chatterjee:2020p6343,Moura:2022p4769,Moura:2022p8041} the energies and properties of charged electronic states are computed by directly approximating the spectral form of forward and backward 1-GF in \cref{eq:freq_GF_spectral_matrix}, which can be written in a non-diagonal matrix representation
\begin{align}
\label{eq:GF_nondiag_matrix}
	\mathbf{G}_{\pm}(\omega) = \mathbf{T}_{\pm}(\omega\mathbf{S}_{\pm}-\mathbf{M}_{\pm})^{-1}\mathbf{T}_{\pm}^{\dagger} \
\end{align}
Similar to $\boldsymbol{\tilde{\Omega}}_{\pm}$ and $\mathbf{\tilde{X}}_{\pm}$ in \cref{eq:freq_GF_spectral_matrix}, the matrices $\mathbf{M}_{\pm}$ and $\mathbf{T}_{\pm}$ in \cref{eq:GF_nondiag_matrix} contain information about vertical charged excitation energies and transition probabilities, respectively, but are expressed in a different (noneigenstate) basis of $(N+1)$- and $(N-1)$-electron configurations ($\ket{\Psi_{+\mu}}$ and $\ket{\Psi_{-\mu}}$).
In the ADC literature, $\mathbf{M}_{\pm}$ and $\mathbf{T}_{\pm}$ are usually called the effective Hamiltonian and effective transition moments matrices, respectively.\cite{Dreuw:2014p82}
The $\mathbf{S}_{\pm}$ matrices in \cref{eq:GF_nondiag_matrix} describe the overlap of $(N+1)$- and $(N-1)$-electron basis states within each set ($S_{\pm \mu \nu} = \braket{\Psi_{\pm\mu}|\Psi_{\pm\nu}}$).
Choosing $\ket{\Psi_{+\mu}}$ and $\ket{\Psi_{-\mu}}$ such that $\mathbf{G}_{+}(\omega)$ and $\mathbf{G}_{-}(\omega)$ do not couple, the $\mathbf{M}_{\pm}$, $\mathbf{T}_{\pm}$, and $\mathbf{S}_{\pm}$ matrices are evaluated up to the order $n$ in perturbation theory
\begin{align}
	\label{eq:M_approx}
	\mathbf{M_\pm} &\approx \mathbf{M^{(0)}_\pm} + \mathbf{M^{(1)}_\pm} + \ldots + \mathbf{M^{(n)}_\pm} \\
	\label{eq:T_approx}
	\mathbf{T_\pm} &\approx \mathbf{T^{(0)}_\pm} + \mathbf{T^{(1)}_\pm} + \ldots + \mathbf{T^{(n)}_\pm} \\
	\label{eq:S_approx}
	\mathbf{S_\pm} &\approx \mathbf{S^{(0)}_\pm} + \mathbf{S^{(1)}_\pm} + \ldots + \mathbf{S^{(n)}_\pm}
\end{align}
corresponding to the $n$th-order non-Dyson ADC approximation (ADC($n$)).
The ADC($n$) charged excitation energies are computed as eigenvalues of the effective Hamiltonian matrices ($\boldsymbol{\Omega}_{\pm}$) 
\begin{align}
	\label{eq:ADC_eig}
	\mathbf{M}_{\pm} \mathbf{Y}_{\pm}=\mathbf{S}_{\pm}\mathbf{Y}_{\pm}\boldsymbol{\Omega}_{\pm}
\end{align}
Combining the  eigenvectors $\textbf{Y}_{\pm}$ with the effective transition moments matrices $\mathbf{T_\pm}$ allows to obtain the ADC($n$) spectroscopic amplitudes $\mathbf{X}_{\pm}$ 
\begin{align}
\label{eq:spec_amp}
    \mathbf{X}_{\pm} = \mathbf{T}_{\pm} \mathbf{S}^{-1/2}_{\pm} \mathbf{Y}_{\pm} 
\end{align}
and transition probabilities (also known as spectroscopic factors)
\begin{align}
	\label{eq:spec_factors}
	P_{\pm\mu} = \sum_{p} |X_{\pm p\mu}|^{2} 
\end{align}
which can be used to compute density of states
\begin{align}
	\label{eq:spec_function}
	A(\omega) &= -\frac{1}{\pi} \mathrm{Im} \left[ \mathrm{Tr} \, \mathbf{G}_{\pm}(\omega) \right]
\end{align}
and simulate photoelectron spectra.
In addition, $\textbf{Y}_{\pm}$ and $\mathbf{X}_{\pm}$ can be used to compute other useful properties of charged excited states (\cref{sec:capabilities_accuracy:properties}), such as electronic density difference, excited-state dipole moments, spin, and Dyson orbitals  
\begin{align}
	\label{eq:Dyson_orb}
	|\phi^{Dyson}_{\pm\mu}\rangle =  \sum_{p} X_{\pm p\mu}  |\phi_{p}\rangle 
\end{align}
that are defined in the basis of reference molecular orbitals ($\phi_{p}$) and are useful for interpreting transitions in photoelectron spectra.

While the Dyson and non-Dyson ADC approaches have been shown to produce similar numerical results at the same level of approximation ($n$),\cite{Trofimov:2005p144115}  the non-Dyson ADC framework allows for more computationally efficient and straightforward simulations of charged excitations by decoupling the forward and backward components of 1-GF and calculating the energies and properties of electron-attached and ionized states independently.
Importantly, the non-Dyson ADC perturbation expansion in \cref{eq:M_approx,eq:T_approx,eq:S_approx} is generated by separating the total Hamiltonian ($H$) into zeroth-order ($H^{(0)}$) and perturbation contributions ($V$), $H = H^{(0)} + V$, and various choices of $H^{(0)}$ (and the corresponding zeroth-order reference wavefunction) are possible for a variety of applications.
In the following, we briefly review three formulations of non-Dyson ADC theory, namely: single-reference ADC for weakly correlated molecular systems (SR-ADC, \cref{sec:theoretical_background:sr_adc}), multireference ADC for molecules with multiconfigurational electronic structure (MR-ADC, \cref{sec:theoretical_background:mr_adc}), and periodic SR-ADC for crystalline materials (\cref{sec:theoretical_background:periodic_adc}).

\subsection{Single-reference ADC}
\label{sec:theoretical_background:sr_adc}

\begin{figure*}[t!]
	\includegraphics[width=0.6\textwidth]{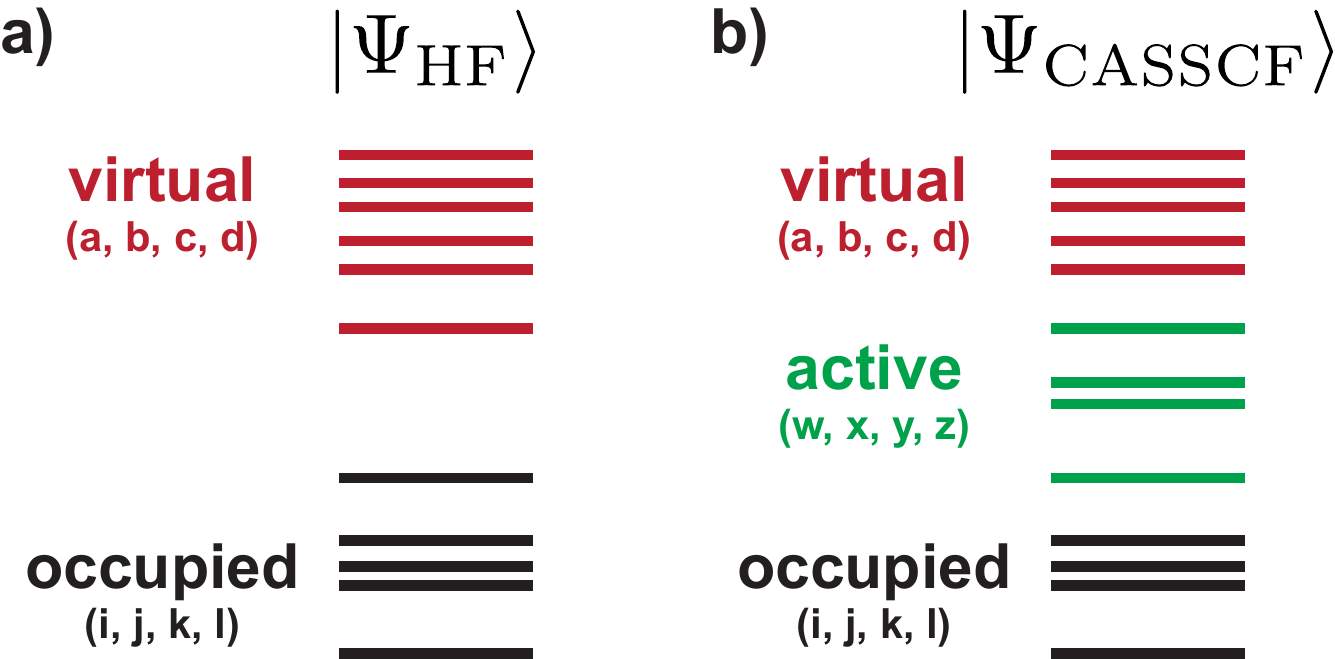} 
	\captionsetup{justification=justified,singlelinecheck=false,font=footnotesize}
	\caption{Schematic diagram representing molecular orbital spaces and their labeling for (a) the Hartree--Fock reference wavefunction in SR-ADC and (b) the CASSCF reference wavefunction in MR-ADC.} 
	\label{fig:orbital_diagram}
\end{figure*}

The single-reference formulation of non-Dyson ADC (SR-ADC)\cite{Schirmer:1998p4734,Schirmer:2001p10621,Thiel:2003p2088,Trofimov:2005p144115,Schneider:2015p144103,Banerjee:2019p224112,Dempwolff:2019p064108,Dempwolff:2020p024113,Dempwolff:2020p024125,Dempwolff:2021p104117,Banerjee:2021p074105} starts by assuming that the ground-state electronic structure of a chemical system can be well described by a Slater determinant wavefunction obtained from a converged Hartree--Fock calculation ($\ket{\Psi_0^N} \approx \ket{\Psi_{\mathrm{HF}}}$).
As depicted in \cref{fig:orbital_diagram}(a), the Slater determinant $\ket{\Psi_{\mathrm{HF}}}$ is represented in a basis of occupied molecular orbitals (labeled as $i,j,k,l, \ldots$), with the remaining virtual orbitals left completely unoccupied (indexed as $a,b,c,d,\ldots$).
Choosing the Fock operator as the zeroth-order Hamiltonian $H^{(0)}$ and incorporating all electron correlation effects in the perturbation operator $V$ defines the order expansions for the ADC matrices in \cref{eq:M_approx,eq:T_approx,eq:S_approx} and the hierarchy of SR-ADC($n$) approximations.

Working equations for the matrix elements of $\mathbf{M}_{\pm}$, $\mathbf{T}_{\pm}$, and $\mathbf{S}_{\pm}$ are derived using one of two alternative theoretical frameworks, namely: (i) the intermediate state representation (ISR) \cite{Schirmer:1991p4647,Mertins:1996p2140,Schirmer:2004p11449} or (ii) the effective Liouvillian theory (EL). \cite{Prasad:1985p1287,Mukherjee:1989p257,Sokolov:2018p204113,Banerjee:2019p224112}
These techniques offer different approaches for incorporating electron correlation effects and treating the coupling between $\textbf{G}_{+}(\omega)$ and $\textbf{G}_{-}(\omega)$, but result in numerically equivalent approximations at the same order in perturbation theory.
We refer the readers to relevant publications for the details behind each approach.\cite{Schirmer:1991p4647,Mertins:1996p2140,Schirmer:2004p11449,Prasad:1985p1287,Mukherjee:1989p257,Sokolov:2018p204113,Banerjee:2019p224112} 
In addition to deriving equations for $\mathbf{M}_{\pm}$, $\mathbf{T}_{\pm}$, and $\mathbf{S}_{\pm}$, both ISR and EL enable calculating density matrices and operator expectation values,\cite{Dempwolff:2020p024113,Dempwolff:2020p024125,Dempwolff:2021p104117,Stahl:2022p044106} which can be used to simulate excited-state properties (\cref{sec:capabilities_accuracy:properties}). 

\begin{figure*}[t!]
	\centering
	\includegraphics[width=1.0\textwidth]{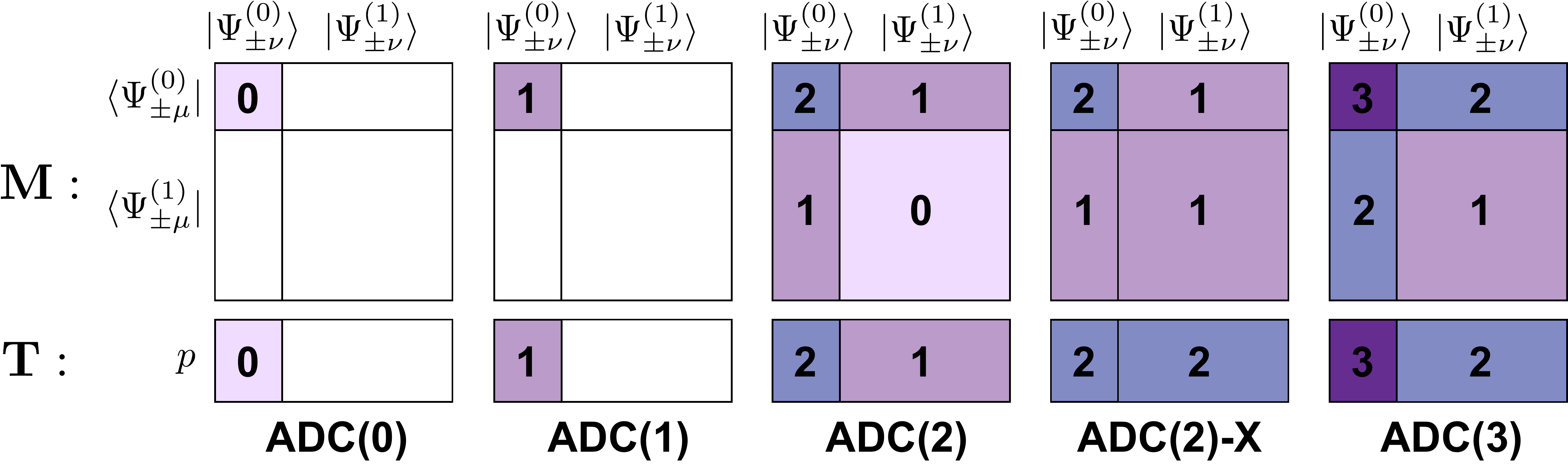}
	\captionsetup{justification=justified,singlelinecheck=false,font=footnotesize}
	\caption{Perturbative structures of the effective Hamiltonian ($\mathbf{M}$) and transition moments ($\mathbf{T}$) matrices in the low-order ADC approximations.
		Numbers denote the perturbation order to which the effective Hamiltonian and transition moments are expanded for each sector.
		Shaded areas indicate non-zero blocks.
	}
	\label{fig:matrix}
\end{figure*}	

\begin{figure*}[t!]
	\includegraphics[width=0.6\textwidth]{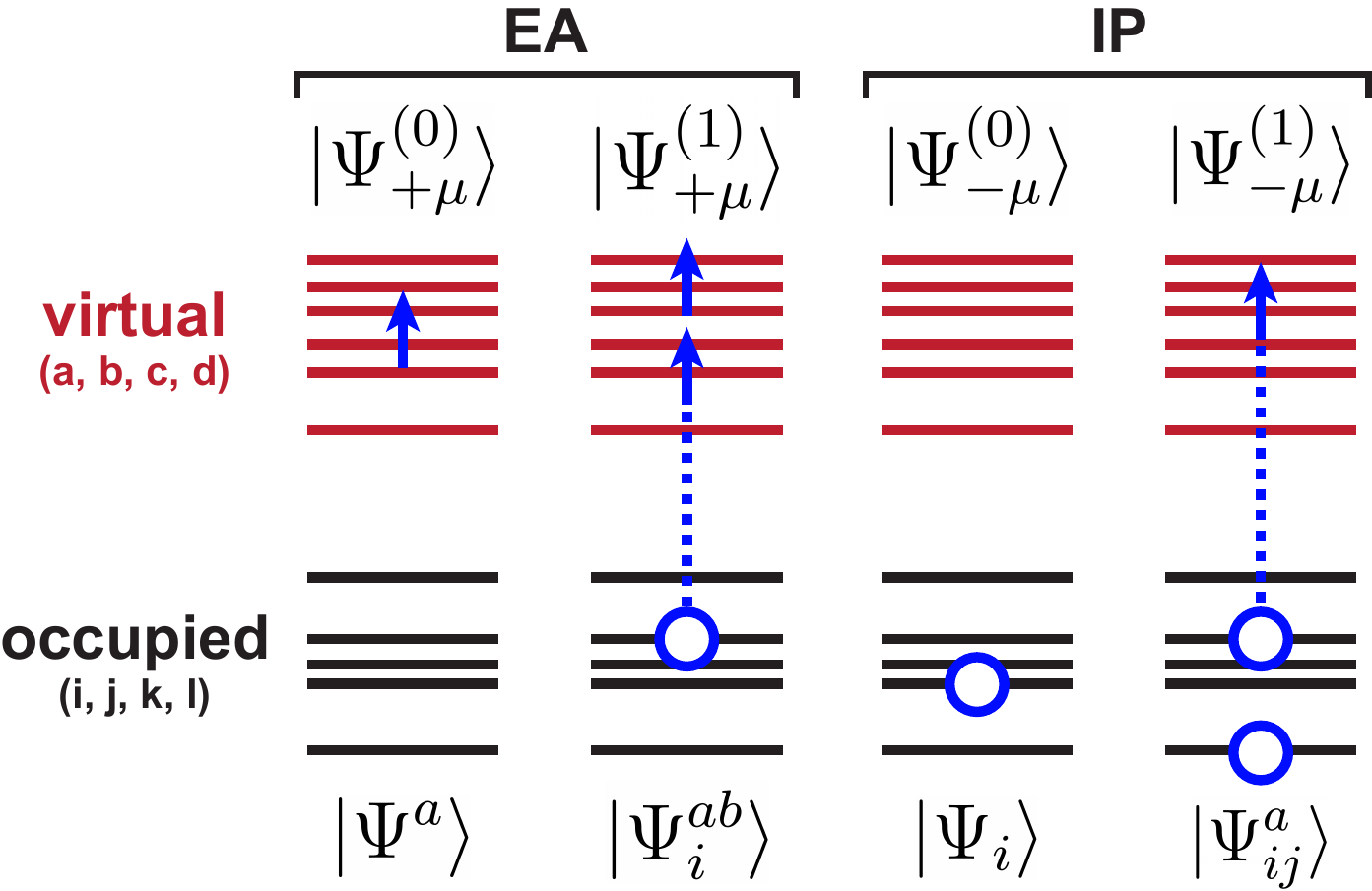}
	\captionsetup{justification=justified,singlelinecheck=false,font=footnotesize}
	\caption{Schematic representation of the electron-attached ($\ket{\Psi^{(k)}_{+\mu}}$) and ionized ($\ket{\Psi^{(k)}_{-\mu}}$) electronic configurations used to represent charged excited states in low-order SR-ADC approximations. 
	An arrow denotes electron attachment, a circle indicates ionization, while a circle connected with an arrow represents a single excitation. 
		\label{fig:sr_ex}}
\end{figure*}

\cref{fig:matrix} shows the structures of $\mathbf{M}_{\pm}$ and $\mathbf{T}_{\pm}$ for the low-order SR-ADC($n$) approximations ($n \le 3$).
Regardless of the approach used to derive working equations, each matrix element can be assigned to one or a pair of excited electronic configurations denoted as $\ket{\Psi^{(k)}_{\pm\mu}}$, where $k$ indicates a perturbation order.
These $(N \pm 1)$-electron  configurations are schematically depicted in \cref{fig:sr_ex}.
The matrices of SR-ADC(0) and SR-ADC(1) approximations are expressed in the basis of zeroth-order states $\ket{\Psi^{(0)}_{+\mu}}$ and $\ket{\Psi^{(0)}_{-\mu}}$ that describe electron attachment to a virtual orbital and ionization of an occupied orbital, respectively (so-called 1p and 1h excitations). 
The first-order states $\ket{\Psi^{(1)}_{\pm\mu}}$ appearing in the equations of higher-order approximations (SR-ADC($n$), $n \ge 2$) describe two-electron processes where attachment or ionization is accompanied by a one-electron excitation (2p-1h or 2h-1p).
In SR-ADC, all basis states $\ket{\Psi^{(0)}_{\pm\mu}}$ are orthogonal to each other with $\mathbf{S}_{\pm} = \mathbf{1}$ at any level of approximation.

Each sector of $\mathbf{M}_{\pm}$ corresponding to a pair of configurations $\bra{\Psi^{(k)}_{\pm\mu}}$ and $\ket{\Psi^{(l)}_{\pm\nu}}$ (\cref{fig:matrix}) is evaluated up to the order $m = n - k - l$, where $n$ is the level of SR-ADC($n$) approximation.
For example, the 1h/1h and 1p/1p blocks calculated with respect to $\bra{\Psi^{(0)}_{\pm\mu}}$ and $\ket{\Psi^{(0)}_{\pm\nu}}$ are expanded up to the second order in SR-ADC(2) and up to the third order in SR-ADC(3), while the 1h/2h-1p and 1p/2p-1h sectors are approximated to order one in SR-ADC(2) and order two in SR-ADC(3).
The $m = n - k - l$ restriction on the perturbation order is violated in the extended SR-ADC(2) approximation (SR-ADC(2)-X) where the 2p-1h/2p-1h and 2h-1p/2h-1p blocks are evaluated up to the first order in perturbation theory.
This approach improves the description of orbital relaxation effects in the simulated excited electronic states, but tends to exhibit larger errors in charged excitation energies compared to SR-ADC(2) due to a lack of error cancellation (\cref{sec:capabilities_accuracy:energies}).
Similar to $\mathbf{M}_{\pm}$, the sectors of $\mathbf{T}_{\pm}$ corresponding to different excited configurations $\ket{\Psi^{(k)}_{\pm\mu}}$ are expanded to order $m = n - k$.

The low-order SR-ADC($n$) ($n \le 3$) methods for electron attachment and ionization have been implemented in the Q-Chem \cite{Epifanovsky:2021p084801} and PySCF\cite{Sun:2020p024109} software packages. 
At each order $n$, SR-ADC($n$) is similar to $n$th-order M\o ller-Plesset perturbation theory (MP$n$)\cite{Moller:1934kp618} in computational cost. 
For a system with $N_{occ}$ occupied and $N_{vir}$ virtual molecular orbitals, the cost of SR-ADC(2) and SR-ADC(3) calculations scales as $\mathcal{O}(N_{occ}^2 N_{vir}^3)$ and $\mathcal{O}(N_{occ}^2 N_{vir}^4)$, respectively.
The SR-ADC(2)-X method has the same computational scaling as SR-ADC(2) for charged excitation energies ($\mathcal{O}(N_{occ}^2 N_{vir}^3)$), but is more expensive for calculating spectroscopic factors ($\mathcal{O}(N_{occ}^2 N_{vir}^4)$).
In practice, these additional costs can be avoided by neglecting expensive terms in $\mathbf{T}_{\pm}$ without significantly affecting the accuracy of SR-ADC(2)-X, resulting in the $\mathcal{O}(N_{occ}^2 N_{vir}^3)$ computational scaling overall.\cite{Banerjee:2019p224112}

\subsection{Multireference ADC}
\label{sec:theoretical_background:mr_adc}

The performance of SR-ADC methods relies on the validity of Hartree--Fock approximation and becomes unreliable when the ground-state electronic structure cannot be accurately represented with a single Slater determinant wavefunction. 
In addition, the low-order SR-ADC($n$) approximations ($n \le 3$) show large errors in transition energies for doubly excited states and do not incorporate three-electron and higher excitations.\cite{Loos:2018p4360}
This significantly reduces the accuracy of SR-ADC calculations for a wide range of important chemical systems, such as transition metal complexes, molecules with unpaired electrons, and highly conjugated organic compounds.

\begin{figure*}[t!]
	\includegraphics[width=1.0\textwidth]{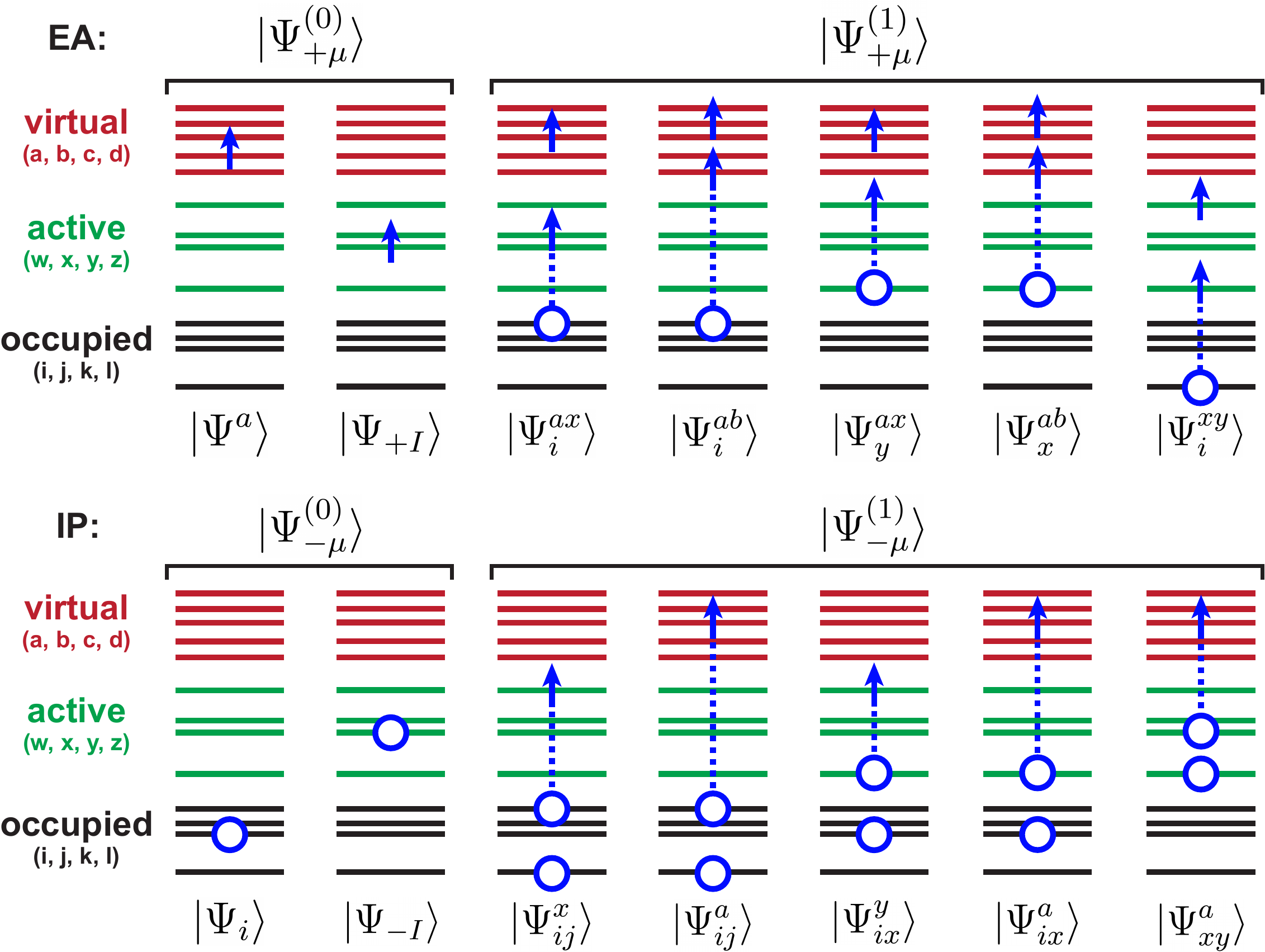}
	\captionsetup{justification=justified,singlelinecheck=false,font=footnotesize}
	\caption{Schematic representation of the electron-attached ($\ket{\Psi^{(k)}_{+\mu}}$) and ionized ($\ket{\Psi^{(k)}_{-\mu}}$) electronic configurations used to represent charged excited states in low-order MR-ADC approximations. 
	An arrow denotes electron attachment, a circle indicates ionization, while a circle connected with an arrow represents a single excitation. 
	}
	\label{fig:mr_exc}
\end{figure*}

The multireference formulation of ADC (MR-ADC) \cite{Sokolov:2018p204113,Chatterjee:2019p5908,Chatterjee:2020p6343,Mazin:2021p6152,Moura:2022p4769,Moura:2022p8041} addresses these problems by describing the ground- and excited-state electronic structure with multiconfigurational wavefunctions.
In MR-ADC, a subset of frontier molecular orbitals is selected to form an active space, as shown in \cref{fig:orbital_diagram}(b).
Next, the molecular orbitals are optimized in a complete active-space self-consistent field (CASSCF) \cite{Werner:1980p2342,Werner:1981p5794,Knowles:1985p259} calculation that computes a multiconfigurational wavefunction for the reference (usually, ground) electronic state ($\ket{\Psi_0^N} \approx \ket{\Psi_{\mathrm{CASSCF}}}$). 
To define the hierarchy of MR-ADC($n$) approximations, the zeroth-order Hamiltonian $H^{(0)}$ is chosen to be the Dyall Hamiltonian,\cite{Dyall:1995p4909} which incorporates all two-electron active-space interactions.
Equations for the $\mathbf{M}_{\pm}$, $\mathbf{T}_{\pm}$, and $\mathbf{S}_{\pm}$ matrix elements (\eqref{eq:M_approx} to \eqref{eq:S_approx}) are derived using a multireference variant of the EL approach.\cite{Sokolov:2018p204113}

Incorporating multireference effects in the active space does not change the perturbative structure of ADC matrices (\cref{fig:matrix}), but introduces new classes of $(N \pm 1)$-electron  configurations $\ket{\Psi^{(0)}_{\pm\mu}}$ and $\ket{\Psi^{(1)}_{\pm\mu}}$ that do not appear in SR-ADC.\cite{Chatterjee:2019p5908,Chatterjee:2020p6343} 
As shown in \cref{fig:mr_exc}, the MR-ADC zeroth-order states $\ket{\Psi^{(0)}_{\pm\mu}}$ describe two kinds of charged excitations: (i) electron attachment or ionization in the active space and (ii) 1p  or 1h excitations in the virtual or occupied orbitals, respectively. 
The active-space charged excitations are represented with complete active-space configuration interaction wavefunctions of the $(N \pm 1)$-electron system ($\ket{\Psi_{\pm I}}$) computed using the reference CASSCF molecular orbitals. 
The 1p and 1h excited configurations ($\ket{\Psi^{a}}$ and $\ket{\Psi_{i}}$) are similar to the 1p and 1h excitations in SR-ADC (\cref{fig:sr_ex}), but are calculated with respect to the CASSCF wavefunction $\ket{\Psi_{\mathrm{CASSCF}}}$ that incorporates multireference effects in the active space.
\cref{fig:mr_exc} shows that the first-order states $\ket{\Psi^{(1)}_{+\mu}}$ and $\ket{\Psi^{(1)}_{-\mu}}$ can be separated into five excitation classes that describe electron attachment or removal accompanied by a single excitation between occupied, active, or virtual orbitals. 
In contrast to SR-ADC, the excited configurations in \cref{fig:mr_exc} are in general non-orthogonal, leading to non-diagonal overlap matrices $\mathbf{S}_{\pm}$.\cite{Chatterjee:2019p5908,Chatterjee:2020p6343} 

The ability to simulate excitations in all molecular orbitals distinguishes MR-ADC from other multireference perturbation theories (MRPT), such as CASPT2 or NEVPT2,\cite{Wolinski:1987p225,Hirao:1992p374,Werner:1996p645,Finley:1998p299,Andersson:1990p5483,Andersson:1992p1218,Angeli:2001p10252,Angeli:2001p297,Angeli:2004p4043} 
that cannot describe electronic transitions outside of active space. 
This makes MR-ADC particularly attractive for simulations of core-level excitations in X-ray spectroscopies (\cref{sec:capabilities_accuracy:core_ionization})\cite{Moura:2022p4769,Moura:2022p8041}  and chemical systems with high density of states.
Additionally, unlike MRPT, MR-ADC offers a straightforward approach to calculate transition properties (e.g., spectroscopic factors, ionization cross sections, etc.) and can compute many excited states without averaging molecular orbitals in the CASSCF calculations. 

The MR-ADC methods for charged excitations have been implemented up to the MR-ADC(2)-X level of theory (\cref{fig:matrix}). \cite{Chatterjee:2019p5908,Chatterjee:2020p6343,Moura:2022p4769,Moura:2022p8041}
In contrast to SR-ADC(2) and SR-ADC(2)-X, the MR-ADC(2)-X method usually shows a better performance compared to MR-ADC(2) for electron attachment and ionization (\cref{sec:capabilities_accuracy:energies,sec:capabilities_accuracy:core_ionization}).
For a fixed active space, the computational scaling of MR-ADC approximations is similar to that of SR-ADC and MRPT methods at each order in perturbation theory. 
Specifically, the computational cost of MR-ADC(2) and MR-ADC(2)-X charged excitation energies scales as \textsc{$\mathcal{O}(N_{occ+act}^2 N_{vir}^3)$}, where $N_{occ+act}$ is the number of occupied and active orbitals, while $N_{vir}$ is the size of virtual space. 
As in SR-ADC, simulating transition properties using MR-ADC(2)-X has a higher computational scaling, which can be lowered back to \textsc{$\mathcal{O}(N_{occ+act}^2 N_{vir}^3)$} by introducing minor approximations.\cite{Chatterjee:2019p5908,Chatterjee:2020p6343}
Increasing the active space size ($N_{act}$) and the number of determinants in the complete active space ($N_{det}$), the computational cost of MR-ADC(2) and MR-ADC(2)-X scales as \textsc{$\mathcal{O}( N_{det} N_{act}^8)$}, similar to CASPT2 or NEVPT2.
As demonstrated in Ref.\@ \citenum{Chatterjee:2020p6343}, this high computational scaling can be decreased to \textsc{$\mathcal{O}( N_{det} N_{act}^6)$} by constructing efficient intermediates without introducing any approximations. 
 
\subsection{Periodic ADC}
\label{sec:theoretical_background:periodic_adc}

Although ADC originated in molecular quantum chemistry, it can be used to simulate the spectroscopic properties of periodic condensed matter systems, such as crystalline solids and low-dimensional materials.
A periodic implementation of Dyson SR-ADC has been developed by Buth and co-workers who applied this approach to ionic crystals and one-dimensional molecular chains. \cite{Buth:2005p195107,Buth:2006p154707,Bezugly:2008p012006} 
Recently, we developed the non-Dyson SR-ADC for periodic systems and demonstrated its applications for a variety of materials from large-gap atomic and ionic solids to small-gap semiconductors. \cite{Banerjee:2022p5337}

In periodic non-Dyson SR-ADC, the Hamiltonian and 1-GF (\cref{eq:freq_GF}) are expressed in a complex-valued basis set of crystalline molecular orbitals \cite{McClain:2017p1209,Sun:2017p164119}
\begin{align}
	\ket{\psi_{p\kc}(\textbf{r})} = \sum_{\mu}c_{\mu p\kc} \phi_{\mu \kc}(\textbf{r})  \label{eq:crys_MO}
\end{align}
constructed as linear combinations of translation-symmetry-adapted Gaussian atomic orbitals 
\begin{align}
	\phi_{\mu\kc}(\textbf{r}) = \sum_{\textbf{T}}e^{i\kc\cdot\textbf{T}}\chi_{\mu}(\textbf{r} - \textbf{T})  \label{eq:crys_AO} 
\end{align}
where $\chi_{\mu}(\textbf{r} - \textbf{T})$ are the atom-centered Gaussian basis functions, $\mathbf{T}$ is a lattice translation vector, and $\kc$ is a crystal momentum vector in the first Brillouin zone. 
Similar to molecular ADC, the forward and backward contributions to 1-GF are written in a non-diagonal form 
\begin{align}
    \textbf{G}_{\pm}(\omega,\kc) = \textbf{T}_{\pm}(\kc)(\omega\textbf{S}_{\pm}(\kc)-\textbf{M}_{\pm}(\kc))^{-1}\textbf{T}_{\pm}^{\dagger}(\kc) \ . \label{eq:G_approx_periodic}
\end{align}
where the ADC matrices acquire dependence on the crystal momentum $\kc$.
Using the approach described in \cref{sec:theoretical_background:sr_adc} and taking care of crystal momentum conservation allows to derive the working equations of periodic SR-ADC($n$) approximations for the matrix elements of $\mathbf{M}_{\pm}(\kc)$, $\mathbf{T}_{\pm}(\kc)$, and $\mathbf{S}_{\pm}(\kc)$.\cite{Banerjee:2022p5337}
The crystal momentum dependence does not change the perturbative structure of ADC matrices (\cref{fig:matrix}) and the nature of electronic configurations that are used to represent charged excited states (\cref{fig:sr_ex}).
As in molecular SR-ADC, in the periodic formulation $\mathbf{S}_{\pm}(\kc) = \mathbf{1}$ at any level of approximation. 

\begin{figure*}[t!]
	\centering
	\captionsetup{justification=justified,singlelinecheck=false,font=footnotesize}
	\subfigure[]{\includegraphics[width=0.7\textwidth]{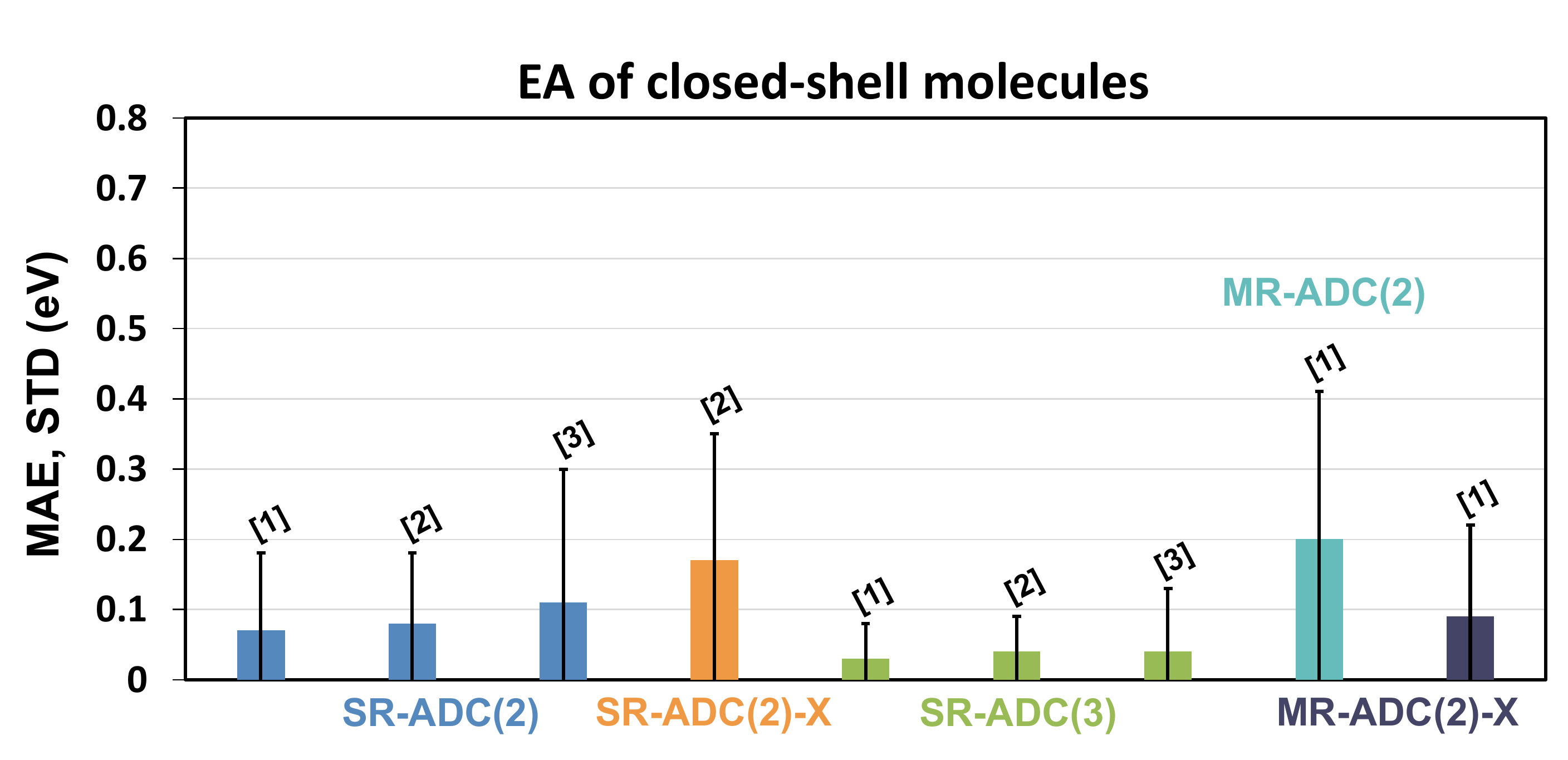}\label{fig:closed_ea}}  
	\subfigure[]{\includegraphics[width=0.7\textwidth]{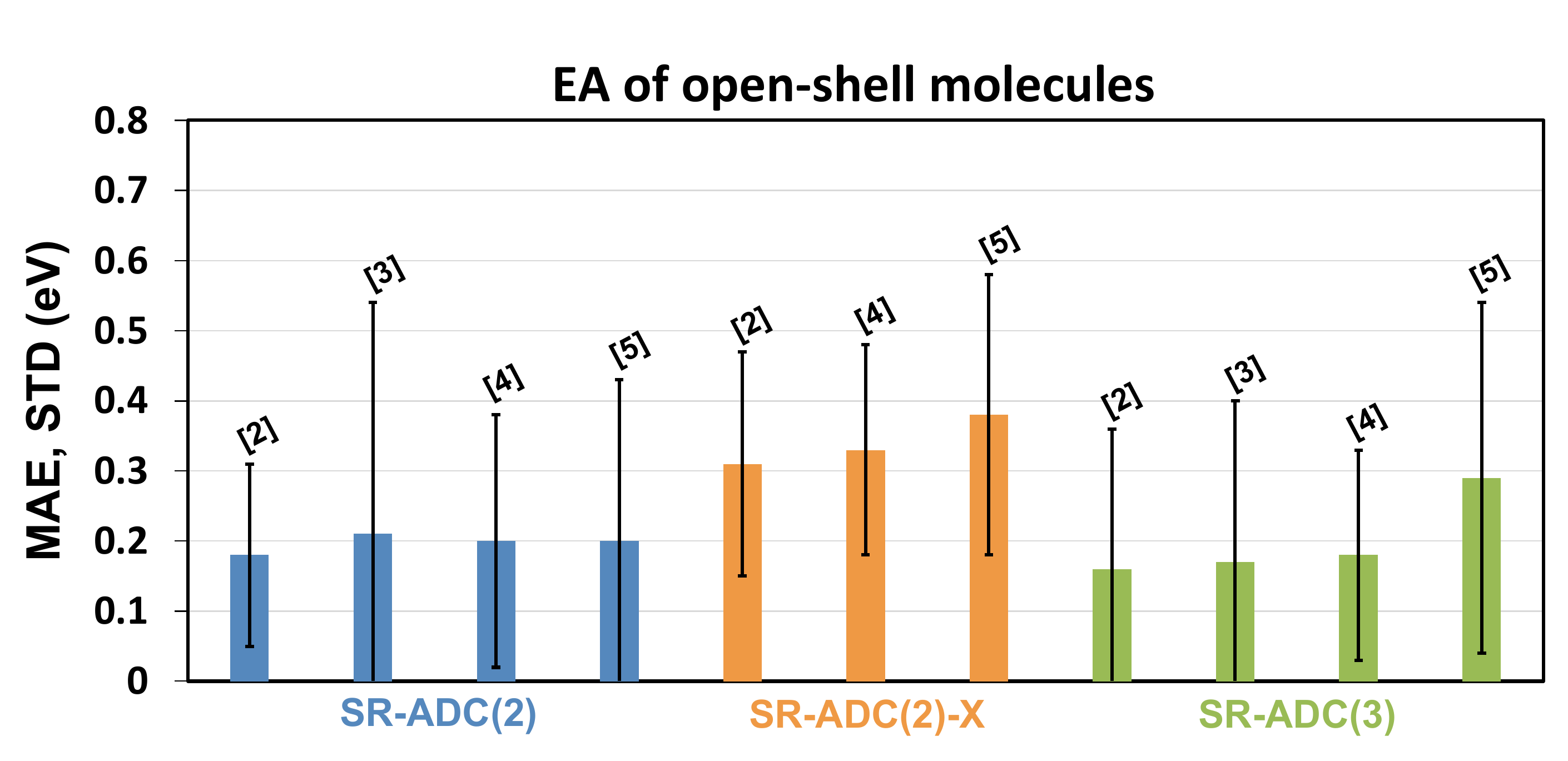}\label{fig:open_ea}}  
	\caption{Mean absolute errors (MAE, eV) and standard deviations (STD, eV) in vertical electron attachment energies of (a) closed-shell and (b) open-shell molecules calculated using different levels of ADC theory. 
		Open-shell calculations were performed using the unrestricted Hartree--Fock reference. 
		MAE are shown as colored boxes, black bars indicate STD multiplied by a factor of two. 
		Data is compiled from different literature sources:
		[1] Ref.\@ \citenum{Chatterjee:2020p6343},
		[2] Ref.\@ \citenum{Banerjee:2019p224112},
		[3] Ref.\@ \citenum{Dempwolff:2021p104117},
		[4] SR-ADC/UHF results for weakly spin-contaminated molecules from Ref.\@ \citenum{Stahl:2022p044106}, and
		[5] SR-ADC/UHF results for strongly spin-contaminated molecules from Ref.\@ \citenum{Stahl:2022p044106}.
	}
	\label{fig:ea_energies}
\end{figure*}

\begin{figure*}[t!]
	\centering
	\captionsetup{justification=justified,singlelinecheck=false,font=footnotesize}
	\subfigure[]{\includegraphics[width=0.7\textwidth]{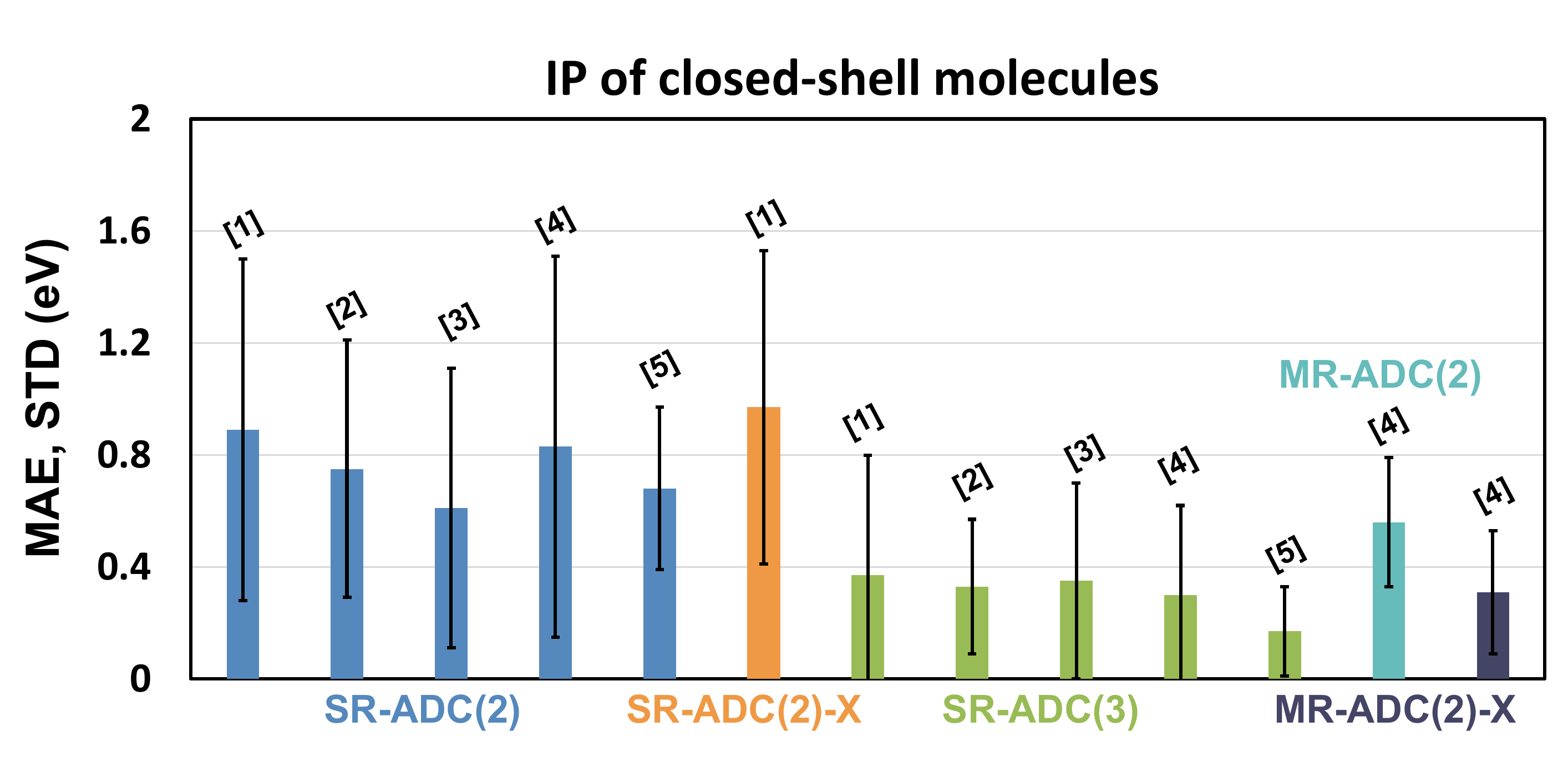}\label{fig:closed_ip}}  
	\subfigure[]{\includegraphics[width=0.7\textwidth]{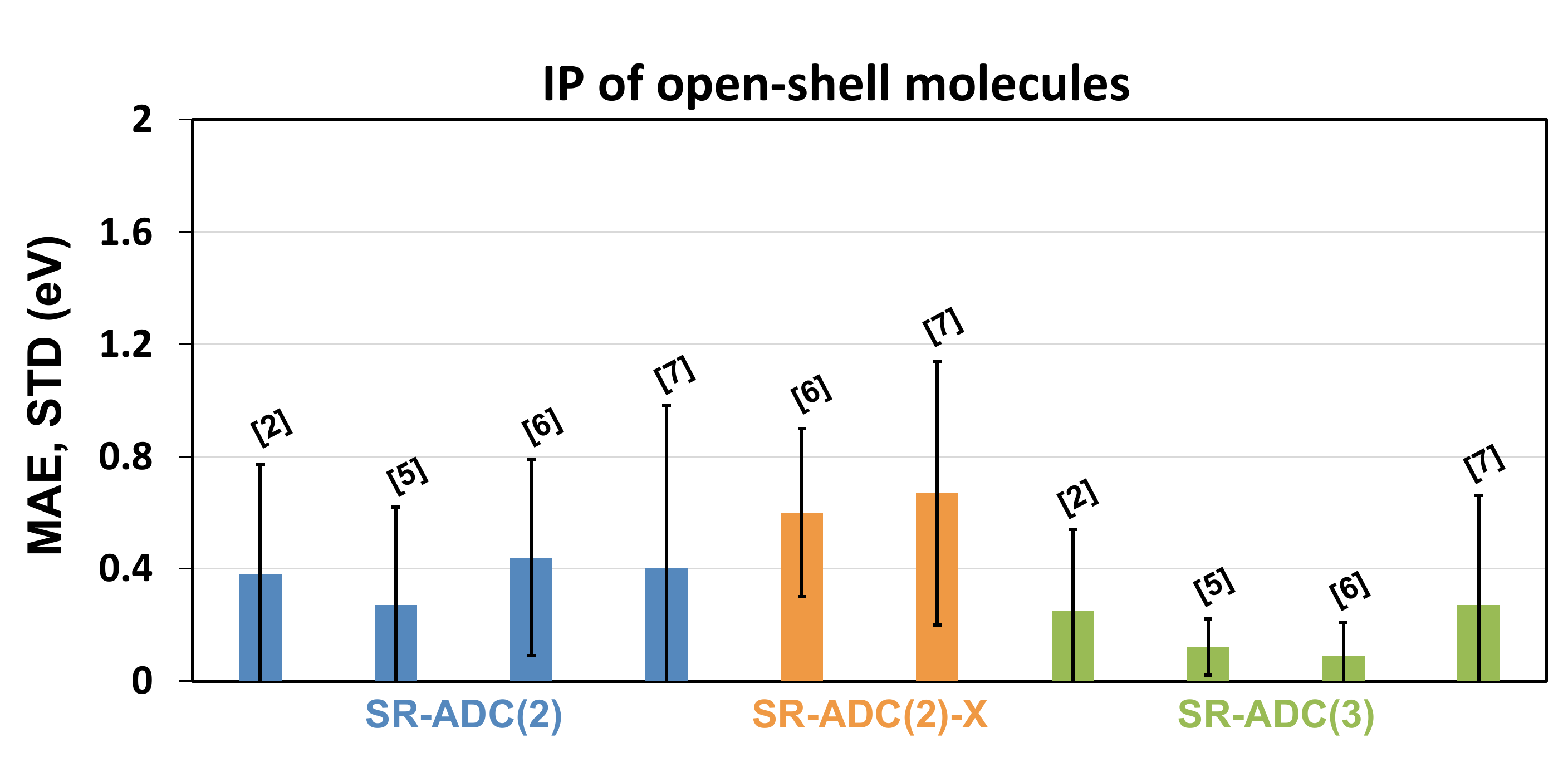}\label{fig:open_ip}}  
	\captionsetup{justification=justified,singlelinecheck=false,font=footnotesize}
	\caption{Mean absolute errors (MAE, eV) and standard deviations (STD, eV) in vertical ionization energies of (a) closed-shell and (b) open-shell molecules calculated using different levels of ADC theory. 
		Open-shell calculations were performed using the unrestricted Hartree--Fock reference. 
		MAE are shown as colored boxes, black bars indicate STD multiplied by a factor of two. 
		Data is compiled from different literature sources:
		[1] Ref.\@ \citenum{Trofimov:2005p144115},
		[2] Ref.\@ \citenum{Dempwolff:2020p024125},
		[3] Ref.\@ \citenum{Dempwolff:2022p054114},
		[4] Ref.\@ \citenum{Chatterjee:2020p6343},
		[5] Ref.\@ \citenum{Dempwolff:2019p064108},
		[6] SR-ADC/UHF results for weakly spin-contaminated molecules from Ref.\@ \citenum{Stahl:2022p044106}, and
		[7] SR-ADC/UHF results for strongly spin-contaminated molecules from Ref.\@ \citenum{Stahl:2022p044106}.
	}
	\label{fig:ip_energies}
\end{figure*}

\begin{figure*}[t!]
	\subfigure[]{\includegraphics[width=0.8\textwidth]{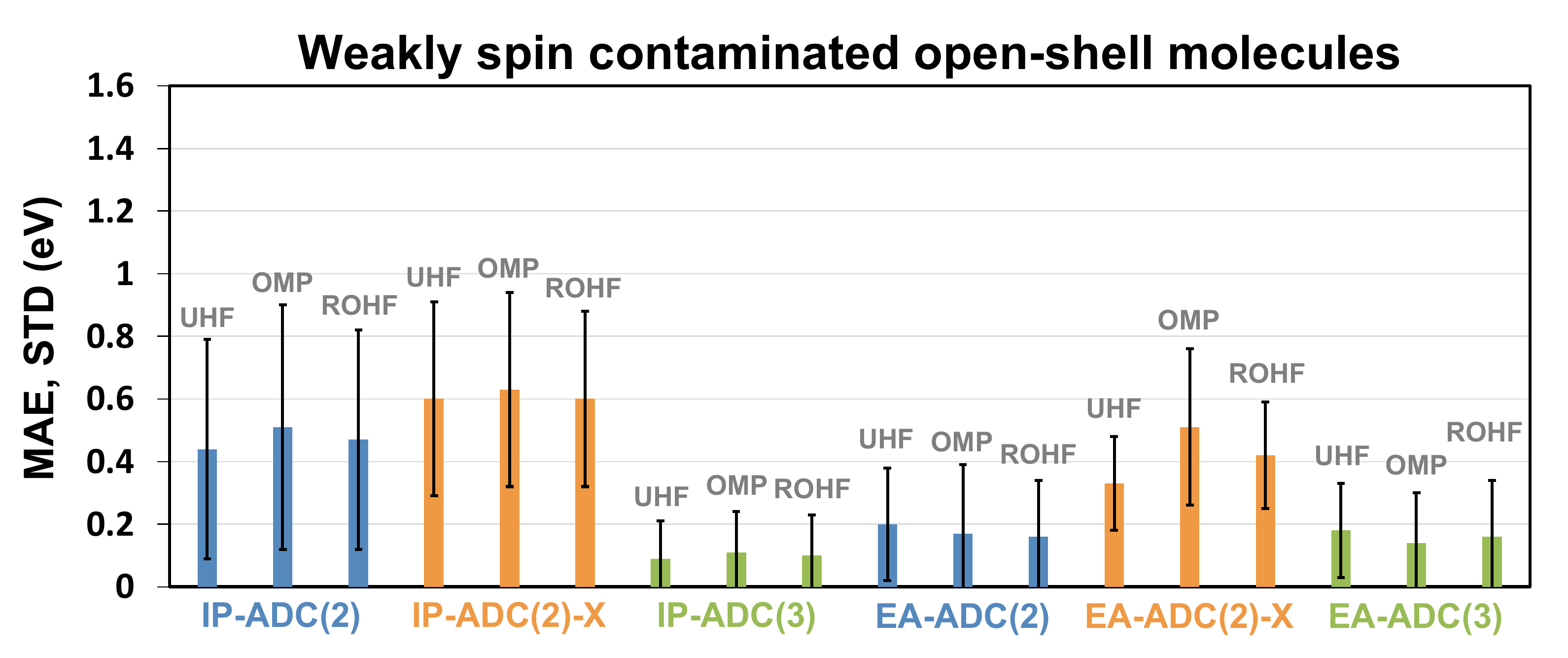}\label{fig:wc_adc}}  
	\subfigure[]{\includegraphics[width=0.8\textwidth]{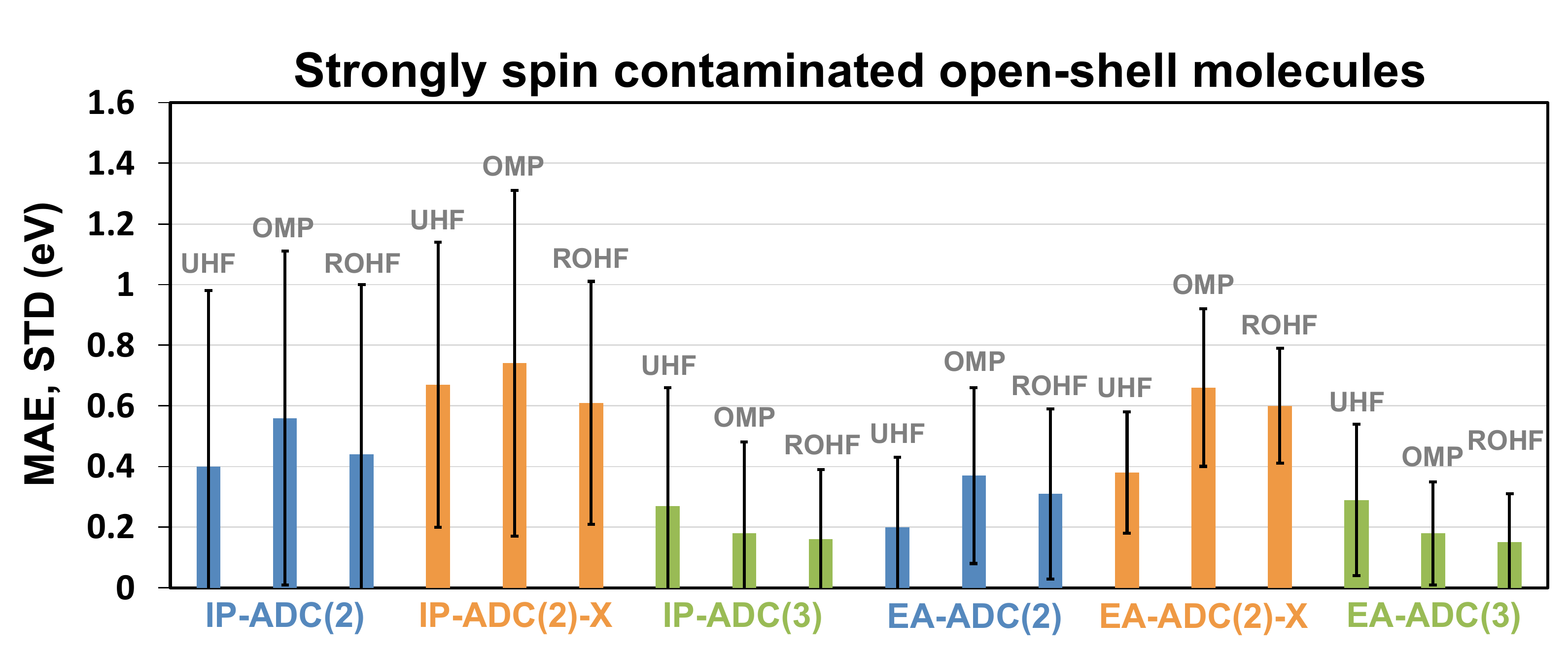}\label{fig:sc_adc}}  
	\captionsetup{justification=justified,singlelinecheck=false,font=footnotesize}
	\caption{Mean absolute errors (MAE, eV) and standard deviations (STD, eV) in the SR-ADC vertical ionization and electron attachment energies for (a) 18 weakly and (b) 22 strongly spin-contaminated molecules with the ground-state UHF spin contamination of $<$ 0.1 and $\geq$ 0.1 a.u., respectively. 
		MAE are shown as colored boxes, black bars indicate STD multiplied by a factor of two. 
		Adapted from Ref.\@ \citenum{Stahl:2022p044106}, with the permission of AIP Publishing.
	}
	\label{fig:sc}
\end{figure*}

The periodic non-Dyson SR-ADC($n$) methods have been implemented in the PySCF package\cite{Sun:2020p024109} up to the third order in perturbation theory ($n \le 3$).\cite{Banerjee:2022p5337}
Taking advantage of the crystalline translational symmetry, the SR-ADC(2) and SR-ADC(2)-X implementations have a $\mathcal{O}(N_{k}^3 N_{occ}^2 N_{vir}^3)$ computational scaling, where $N_{occ}$ and $N_{vir}$ are the numbers of occupied and virtual orbitals per unit cell and $N_{k}$ is the number of $k$-points sampled in the first Brillouin zone.
The computational cost of periodic SR-ADC(3) is dominated by the non-iterative calculation of ground-state wavefunction parameters, which scales as $\mathcal{O}(N_{k}^4 N_{occ}^2 N_{vir}^4)$.
For the remaining steps of SR-ADC(3) algorithm, the computational scaling does not exceed $\mathcal{O}(N_{k}^3 N_{occ}^2 N_{vir}^3)$.
 
\section{Capabilities and Accuracy of ADC for Charged Excitations}
\label{sec:capabilities_accuracy}

\subsection{Electron affinities and ionization energies}
\label{sec:capabilities_accuracy:energies}

The ADC calculations of charged excitations provide electron attachment and ionization energies that are computed by diagonalizing the effective Hamiltonian matrix in \cref{eq:ADC_eig}.
\cref{fig:ea_energies,fig:ip_energies} show the error statistics for simulating vertical electron attachment and ionization energies of closed- and open-shell molecules using various levels of ADC theory, compiled from different benchmark studies.
\cite{Trofimov:2005p144115,Banerjee:2019p224112,Chatterjee:2020p6343,Dempwolff:2019p064108,Dempwolff:2020p024125,Dempwolff:2022p054114,Dempwolff:2021p104117,Stahl:2022p044106}
For closed-shell systems, increasing the level of theory from SR-ADC(2) to SR-ADC(3) reduces the mean absolute error (MAE) in vertical charged excitation energies by a factor of two (\cref{fig:closed_ea,fig:closed_ip}).
SR-ADC(3) shows the MAE of $\sim$ 0.05 eV in electron affinities and  $\sim$ 0.35 eV in ionization energies. 
The performance of SR-ADC(2)-X is similar to SR-ADC(2) for ionization energies and is slightly worse than SR-ADC(2) for electron affinities.

\cref{fig:open_ea,fig:open_ip} demonstrate that the accuracy of SR-ADC calculations is different for molecules with unpaired electrons in the ground electronic state. 
These calculations, performed using the unrestricted Hartree--Fock (UHF) reference wavefunctions, present new challenges, such as accurate description of electronic spin states and increased importance of electron correlation effects.\cite{Stahl:2022p044106}
Relative to closed-shell benchmark results, the average errors in vertical electron affinities of open-shell molecules increase two-fold for SR-ADC(2) and SR-ADC(2)-X and by more than three-fold for SR-ADC(3) (\cref{fig:open_ea}).
Notably, opposite trend is observed for the ionization energies of open-shell systems (\cref{fig:open_ip}) where the SR-ADC methods show a two-fold decrease in MAE due to fortuitous error cancellation.

In a recent study, Stahl et al.\@ investigated the effect of spin contamination in UHF reference wavefunction ($\Delta S^2$) on the accuracy of open-shell SR-ADC calculations. \cite{Stahl:2022p044106}
Their results demonstrated that the charged excitation energies of SR-ADC(2) and SR-ADC(2)-X are relatively insensitive to spin contamination, while SR-ADC(3) is significantly less accurate for molecules with strongly spin-contaminated UHF references ($\Delta S^2$ $>$ 0.1 a.u.)\@ where the errors in excitation energies increase by at least 60\% (\cref{fig:sc}). 
Combining SR-ADC(3) with restricted open-shell Hartree--Fock (ROHF) or orbital-optimized M\o ller--Plesset (OMP) reference wavefunctions improves performance for strongly spin-contaminated systems.
As discussed in \cref{sec:capabilities_accuracy:properties}, this improvement correlates with a decrease in the ground- and excited-state spin contamination when using the ROHF or OMP reference orbitals.

The accuracy of MR-ADC methods for charged excitation energies has been benchmarked in Refs.\@ \citenum{Chatterjee:2019p5908} and \citenum{Chatterjee:2020p6343} for a variety of small closed-shell molecules.
\cref{fig:closed_ea,fig:closed_ip} show the MR-ADC benchmark data for equilibrium molecular geometries where the ground-state wavefunction is dominated by a single Slater determinant.
In this case, MR-ADC(2)-X is consistently more accurate than MR-ADC(2) with MAE ranging between that of SR-ADC(2) and SR-ADC(3). 
Refs.\@ \citenum{Chatterjee:2019p5908} and \citenum{Chatterjee:2020p6343} also report benchmark results for molecules with multiconfigurational electronic structures.
At stretched molecular geometries, the MAE of MR-ADC(2)-X in vertical ionization energies (0.25 eV) is more than ten times smaller than that of SR-ADC(3) (3.66 eV).
For the \ce{C2} molecule with a multireference ground state, SR-ADC(3) overestimates the first electron affinity by $\sim$ 0.9 eV while MR-ADC(2)-X shows an error of $\sim$ 0.25 eV, relative to accurate reference results from selected configuration interaction.\cite{Chatterjee:2020p6343}

\subsection{Core ionization energies}
\label{sec:capabilities_accuracy:core_ionization}

\begin{figure*}[t!]
	\centering
	\includegraphics[width=0.6\textwidth]{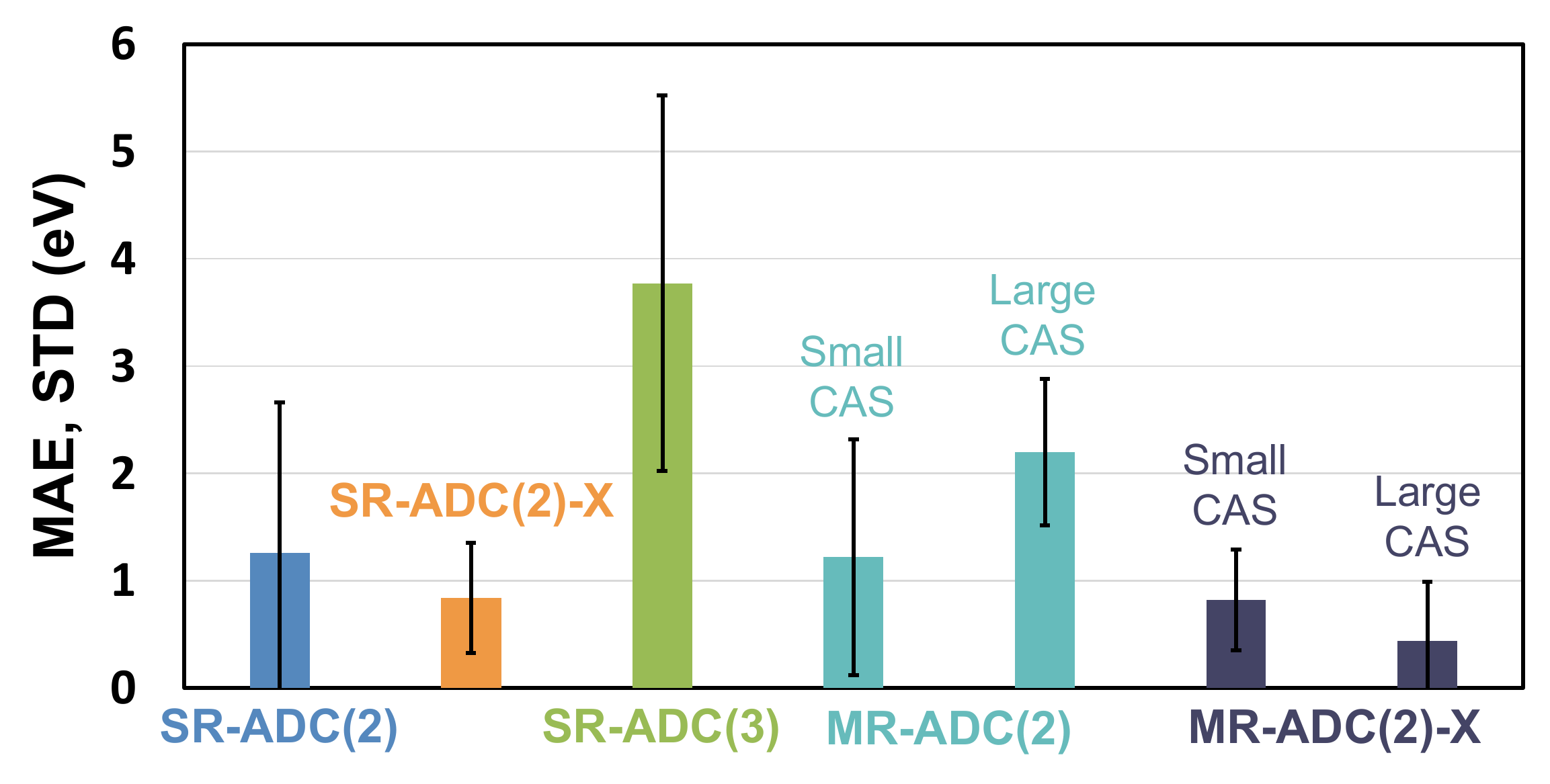}
	\captionsetup{justification=justified,singlelinecheck=false,font=footnotesize}
	\caption{Mean absolute errors (MAE, eV) and standard deviations (STD, eV) in the K-edge core ionization energies of small molecules computed using the SR-ADC and MR-ADC methods, relative to reference EOM-CCSDT results.\cite{Liu:2019p1642}
		MAE are shown as colored boxes, black bars indicate STD multiplied by a factor of two. 
		Adapted from Ref.\@ \citenum{Moura:2022p8041} with permission from the Royal Society of Chemistry.}
	\label{fig:adc_core_ionization}
\end{figure*}

\begin{figure*}[t!]
	\centering
	\includegraphics[width=0.6\textwidth]{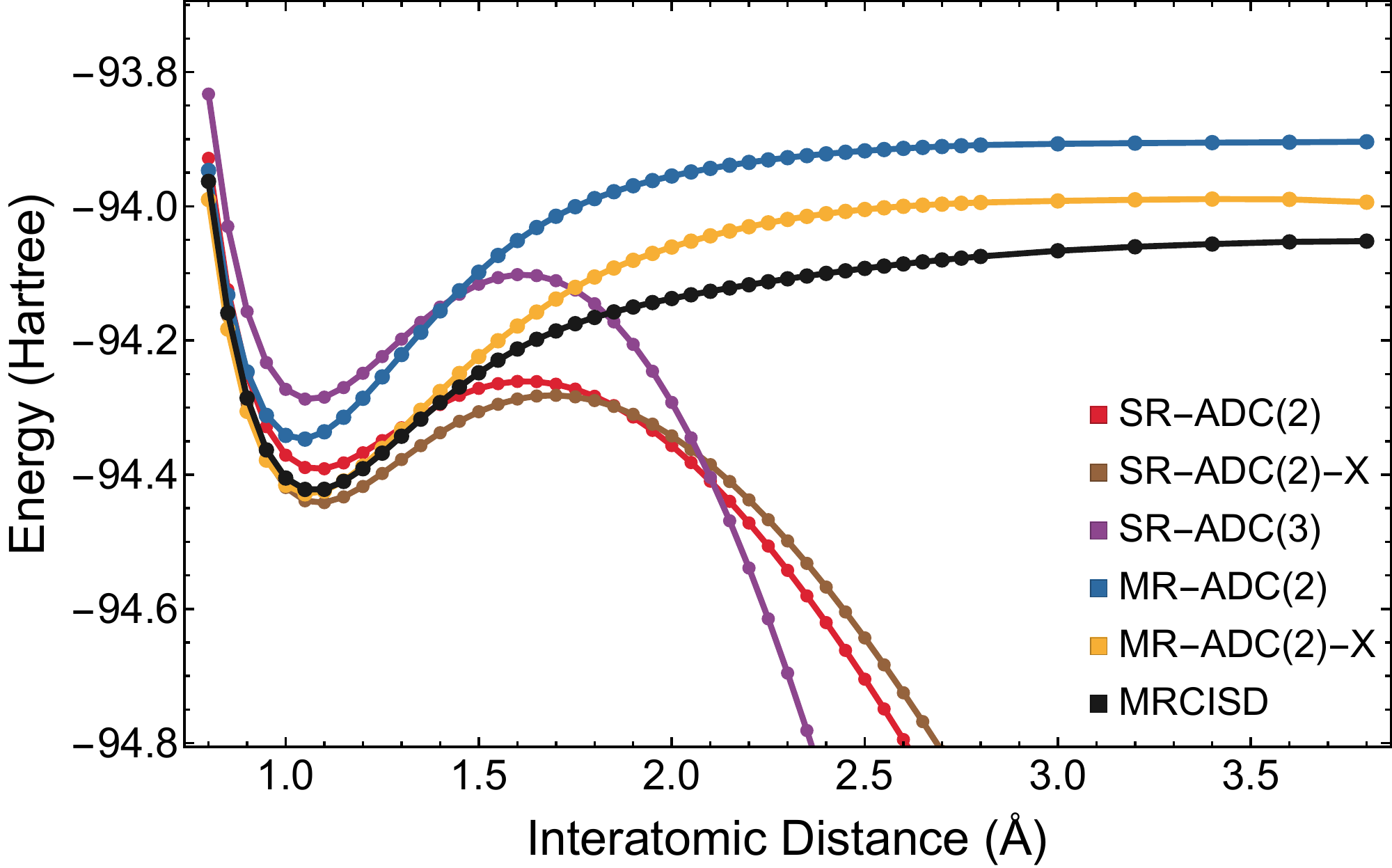}
	\captionsetup{justification=justified,singlelinecheck=false,font=footnotesize}
	\caption{Potential energy curves for the K-edge core-ionized state of \ce{N2} computed using the SR-ADC, MR-ADC, and MRCISD methods. 
		Reproduced from Ref. \citenum{Moura:2022p8041} with permission from the Royal Society of Chemistry.
	}
	\label{fig:N2_PES}
\end{figure*}

In addition to charged excitations in valence orbitals, ADC methods can be used to calculate the energies and properties of core-ionized states that are probed in X-ray photoelectron spectroscopy experiments (XPS).\cite{Chastain:1992p221}
These calculations require special numerical techniques to access core-level excited states that are deeply buried in the  eigenstate spectrum of ADC effective Hamiltonian (\cref{eq:ADC_eig}).
One such technique, termed core-valence separation (CVS) approximation, allows to compute high-energy core-excited states by neglecting their interaction with low-lying states originating from transitions in valence orbitals. \cite{Cederbaum:1980p206,Barth:1981p1038,Koppel:p1997p4415,Trofimov:2000p483,Wenzel:2014p4583,Wenzel:2014p1900,Wenzel:2015p214104,Coriani:2015p181103,Vidal:2019p3117,Liu:2019p1642,HelmichParis:2021pe26559,Thielen:2021p154108} 
The CVS approximation has been widely used in the ADC calculations of core ionization energies,\cite{Angonoa:1987p6789,Dobrodey:2000p7336,Schirmer:2001p10621,Dobrodey:2002p022501,Dobrodey:2002p3533,Dobrodey:2002p165103,Thiel:2003p2088,Plekan:2008p360,Ivanova:2009p545,Holland:2019p224303,Herbst:2020p054114} with some studies exploring alternative variants of this technique.\cite{Banerjee:2019p224112,Moura:2022p4769,Moura:2022p8041}
 
Several SR-ADC studies of core-ionized states have been reported. 
Angonoa et al.\@ developed the first implementation of Dyson SR-ADC(4) and demonstrated its applications to K-shell (1s) ionization in \ce{N2} and CO.\cite{Angonoa:1987p6789}
Schirmer and co-workers presented non-Dyson SR-ADC(4) and showed that this method is equivalent to Dyson SR-ADC(4) within the CVS approximation.\cite{Schirmer:2001p10621,Thiel:2003p2088}
The SR-ADC(4) method has been used to simulate core-ionized states in a variety of small molecules ranging from diatomics to DNA nucleobases.\cite{Thiel:2003p2088,Dobrodey:2000p7336,Dobrodey:2002p022501,Dobrodey:2002p3533,Dobrodey:2002p165103,Plekan:2008p360,Ivanova:2009p545,Holland:2019p224303}
In addition to CVS, a Green's function implementation of SR-ADC(3) has been used to compute the K-, L-, and M-shell ionization energies of a Zn atom.\cite{Banerjee:2019p224112} 

\begin{figure}[t!]
	\centering
	\captionsetup{justification=justified,singlelinecheck=false,font=footnotesize}
	\includegraphics[width=0.4\textwidth]{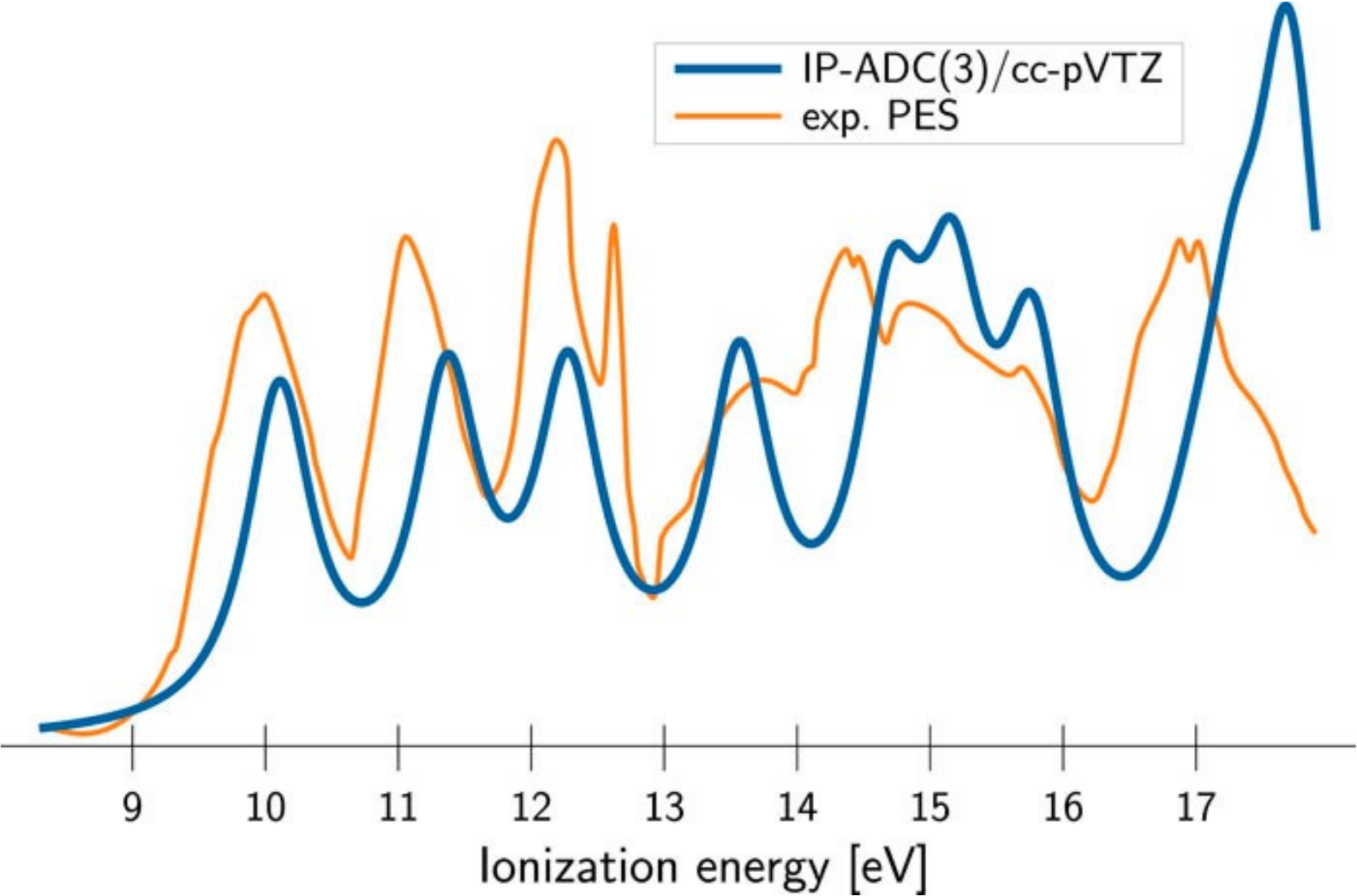}
	\caption{Photoelectron spectrum of the glycine molecule computed using SR-ADC(3) and compared to the experimental results. \cite{Cannington:1983p139}
		Reprinted from Ref.\@ \citenum{Dempwolff:2019p064108}, with the permission of AIP Publishing.
	}
	\label{fig:pes_glycine}
\end{figure}

\begin{figure}[t!]
	\centering
	\captionsetup{justification=justified,singlelinecheck=false,font=footnotesize}
	\subfigure[]{\includegraphics[width=0.4\textwidth]{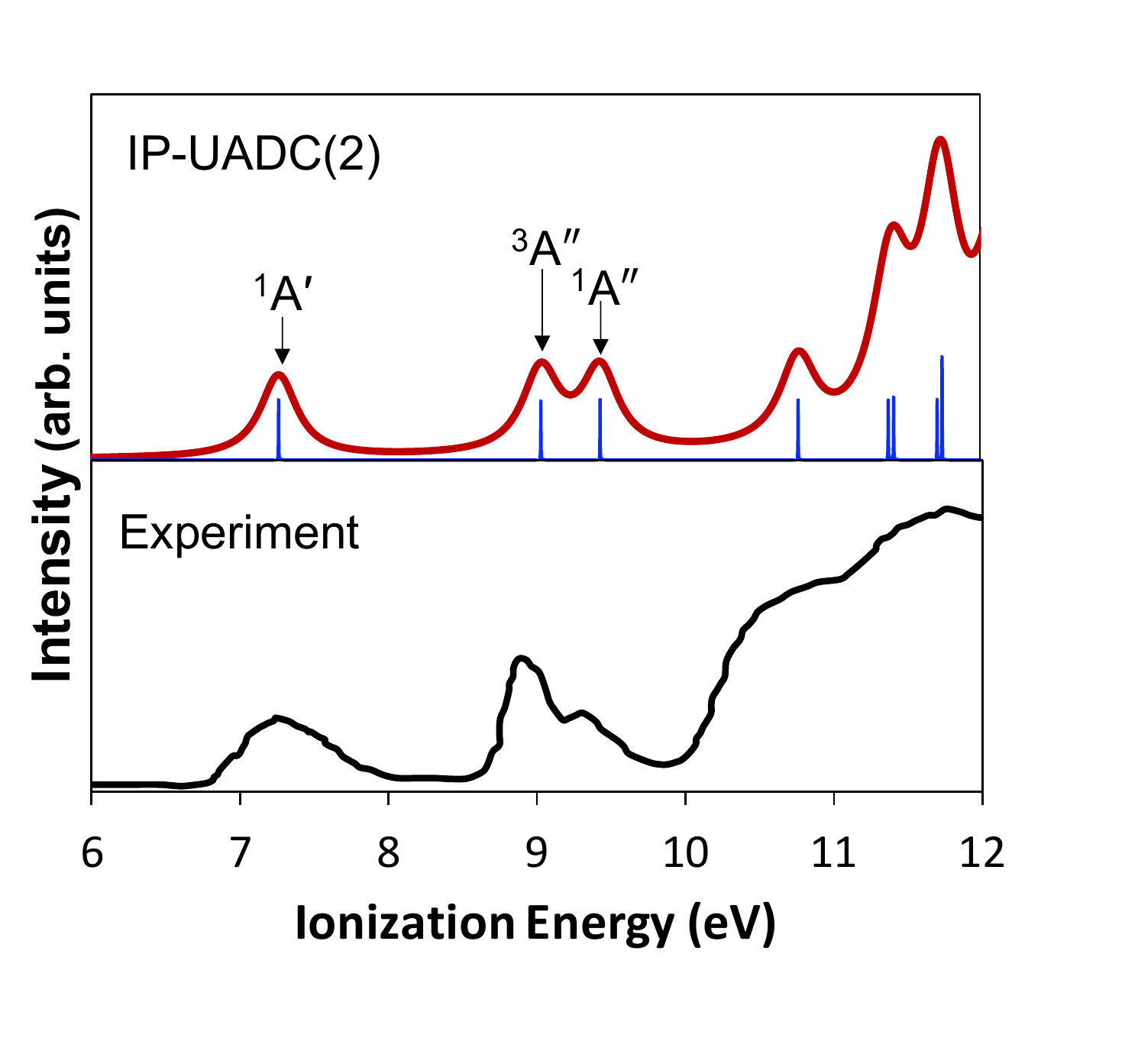}\label{fig:adc2_spectra}}  \qquad
	\subfigure[]{\includegraphics[width=0.4\textwidth]{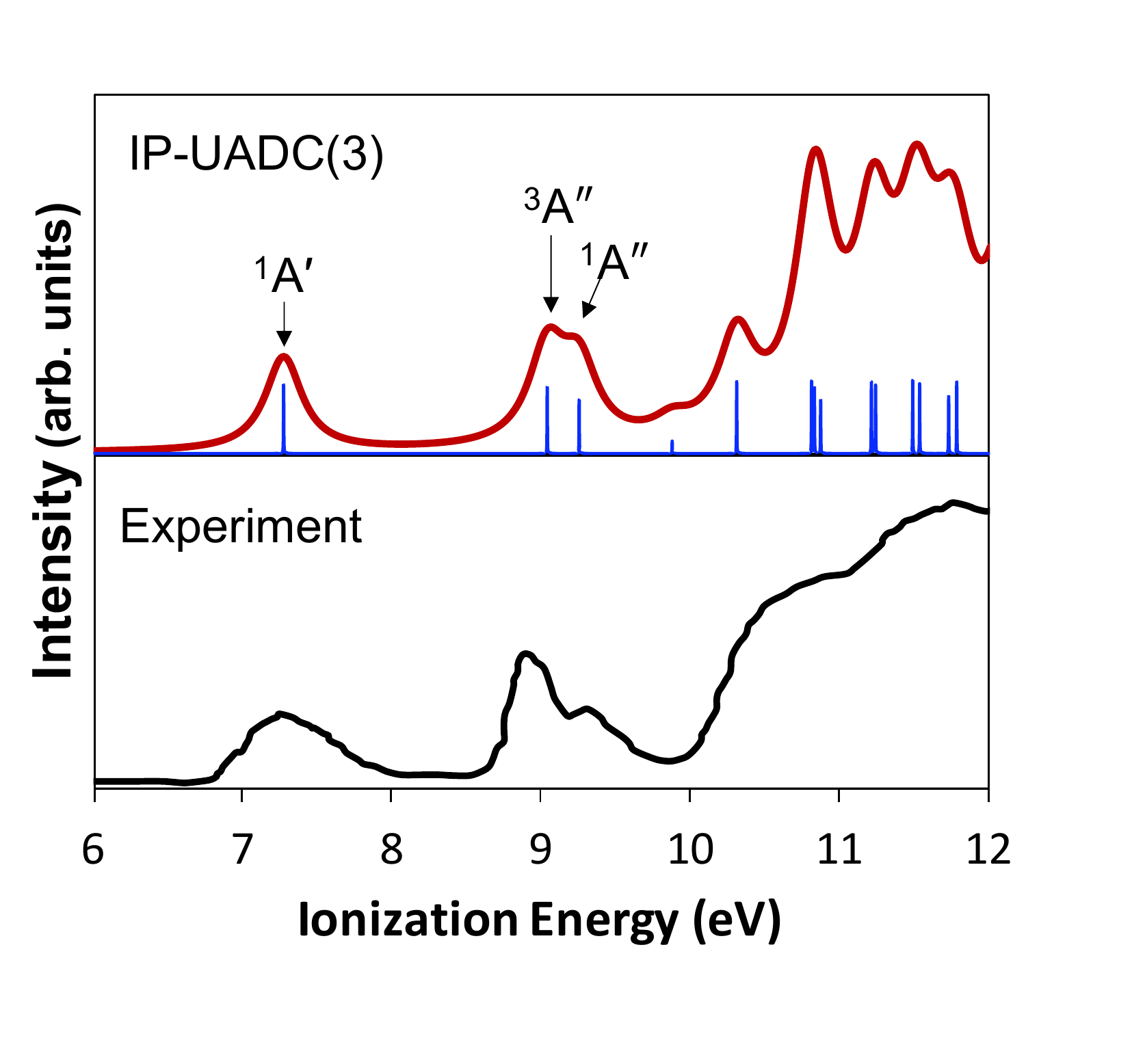}\label{fig:adc3_spectra}}  
	\caption{Photoelectron spectrum of the TEMPO radical computed using (a) SR-ADC(2) and (b) SR-ADC(3) and compared to the experimental results.
		The simulated spectra were shifted by (a) 1.02 and (b) $-$0.3 eV to reproduce the position of first peak in the experimental data. \cite{Kubala:2013p2033} 		
		Reprinted from Ref.\@ \citenum{Banerjee:2021p074105}, with the permission of AIP Publishing.
	}
	\label{fig:pes_tempo}
\end{figure} 

\begin{figure}[t!]
	\centering
	\captionsetup{justification=justified,singlelinecheck=false,font=footnotesize}
	\includegraphics[width=0.4\textwidth]{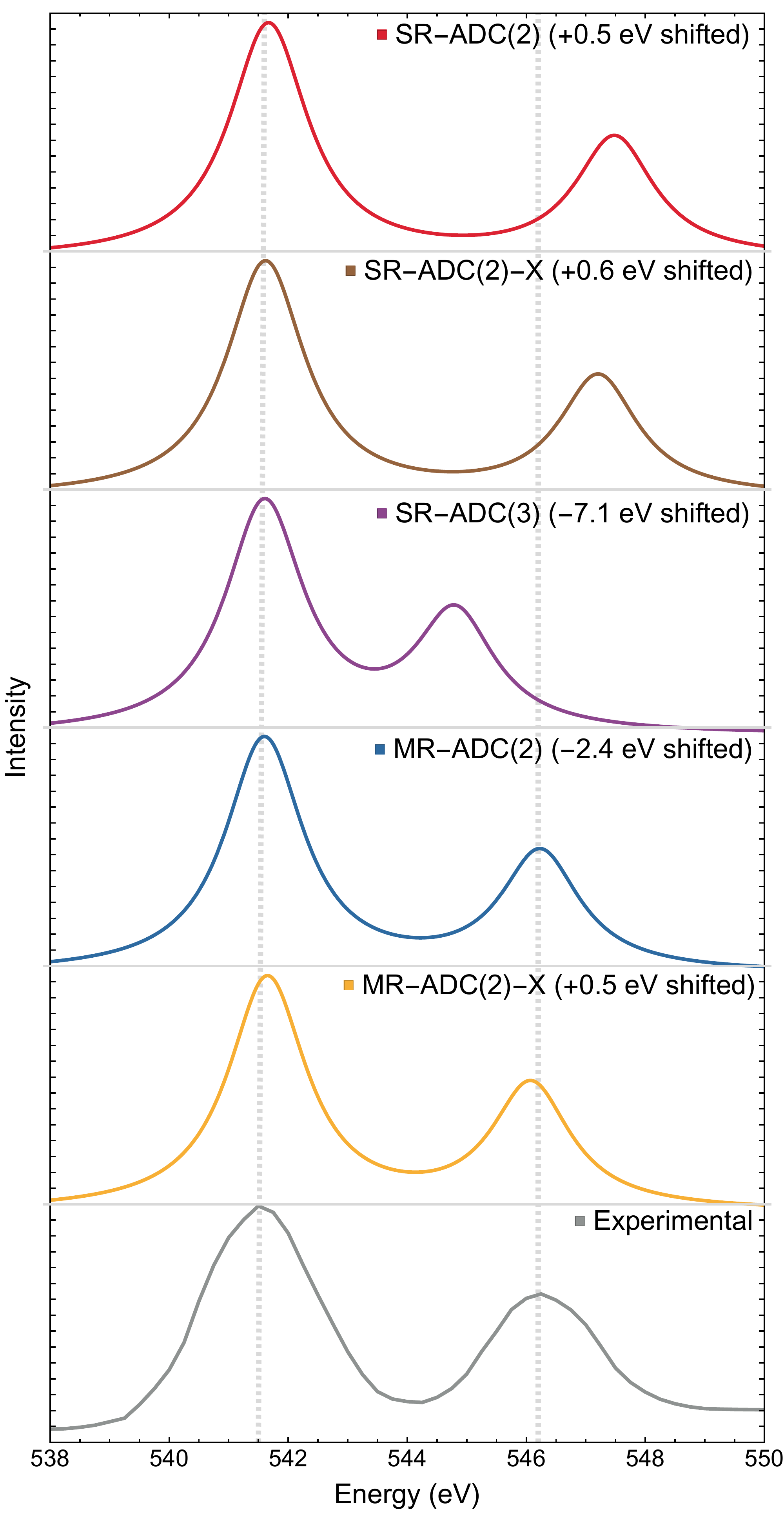}
	\caption{Oxygen K-edge X-ray photoelectron spectrum of ozone computed using the SR-ADC and MR-ADC methods compared to the experimental spectrum from Ref.\@ \citenum{Banna:1977p213}.
		The simulated spectra used a 0.8 eV broadening parameter and were shifted to align with the first peak of the experimental spectrum. 
		Reproduced from Ref.\@ \citenum{Moura:2022p8041} with permission from the Royal Society of Chemistry.
	}
	\label{fig:pes_ozone}
\end{figure} 

Very recently, Refs. \citenum{Moura:2022p4769} and \citenum{Moura:2022p8041} reported a CVS implementation of MR-ADC(2) and MR-ADC(2)-X for simulating core ionization of molecules with multiconfigurational electronic structure.
In contrast to conventional multireference theories, combining MR-ADC with CVS allows to directly access core-ionized states without including core orbitals in the active space.
\cref{fig:adc_core_ionization} shows the error statistics for core ionization energies of small molecules at equilibrium geometries calculated using selected SR- and MR-ADC methods, relative to accurate reference data from EOM-CCSDT.\cite{Liu:2019p1642}
The best results are demonstrated by SR-ADC(2)-X and MR-ADC(2)-X methods, which show errors of less than 1 eV in core ionization energies.
When using a small complete active space (CAS), the performance of MR-ADC methods is very similar to that of SR-ADC.
Increasing the size of CAS shifts the balance of error cancellation leading to larger MAE for MR-ADC(2) and smaller MAE for MR-ADC(2)-X.

The MR-ADC methods can be used to compute potential energy surfaces (PES) of core-ionized states away from equilibrium geometries, which are useful for interpreting the results of time-resolved XPS experiments.
\cref{fig:N2_PES} demonstrates that MR-ADC(2) and MR-ADC(2)-X correctly predict the PES shape for the K-shell-ionized state of \ce{N2}, in a good agreement with reference MRCISD calculations, while the core-ionized PES computed using SR-ADC diverge away from the equilibrium region.\cite{Moura:2022p4769,Moura:2022p8041}

\subsection{Photoelectron spectra}
\label{sec:capabilities_accuracy:spectra}

The ADC methods allow to efficiently simulate photoelectron spectra by calculating the density of states in \cref{eq:spec_function}.
Several SR- and MR-ADC computations of molecular UV and X-ray photoelectron spectra have been reported. \cite{Dobrodey:2000p7336,Dobrodey:2002p022501,Dobrodey:2002p3533,Dobrodey:2002p165103,Plekan:2008p360,Ivanova:2009p545,Holland:2019p224303,Dempwolff:2019p064108,Dempwolff:2020p024113,Banerjee:2021p074105,Moura:2022p4769,Moura:2022p8041,Chatterjee:2019p5908} 

\cref{fig:pes_glycine} compares the experimental photoelectron spectrum of glycine molecule with the results of SR-ADC(3) calculations performed by Dempwolff and co-workers.\cite{Dempwolff:2019p064108}
Although the simulated spectrum does not incorporate vibrational effects, it reproduces the key features of the experimental spectrum quite well. 
\cref{fig:pes_tempo} shows the results of SR-ADC(2) and SR-ADC(3) calculations for the open-shell TEMPO radical from Ref.\@ \citenum{Banerjee:2021p074105}.
The computed spectra were shifted to reproduce the position of first peak in the experimental data.
Apart from this uniform shift, the SR-ADC(2) and SR-ADC(3) photoelectron spectra are in a good agreement with the experimental results.
The SR-ADC(3) calculations show smaller errors in ionization energies and provide a better description of photoelectron spectrum beyond 10 eV.

Examples of X-ray photoelectron spectra simulated using SR- and MR-ADC are shown in \cref{fig:pes_ozone} for the ozone molecule (\ce{O3}).\cite{Moura:2022p4769,Moura:2022p8041}
Although all ADC methods correctly predict the ordering and relative intensities of two peaks corresponding to the K-shell ionization of terminal and central oxygen atoms, the SR-ADC methods show large ($>$ 1 eV) errors in peak spacing relative to the experimental results.\cite{Banna:1977p213}.
These errors originate from the multiconfigurational nature of \ce{O3} ground-state electronic structure, which presents challenges for single-reference theories such as SR-ADC.\cite{Hayes.1971,Hay.1975,Laidig.1981,Schmidt.1998,Kalemos.2008,Musial.2009,Miliordos.2013,Takeshita.2015}
The X-ray photoelectron spectra simulated using MR-ADC(2) and MR-ADC(2)-X are in a good agreement with experimental data showing less than 0.3 eV errors in peak spacing.

\subsection{Properties of periodic systems}
\label{sec:capabilities_accuracy:solids}

\begin{figure*}[t!]
	\captionsetup{justification=justified,singlelinecheck=false,font=footnotesize}	\subfigure[]{\includegraphics[width=0.45\textwidth]{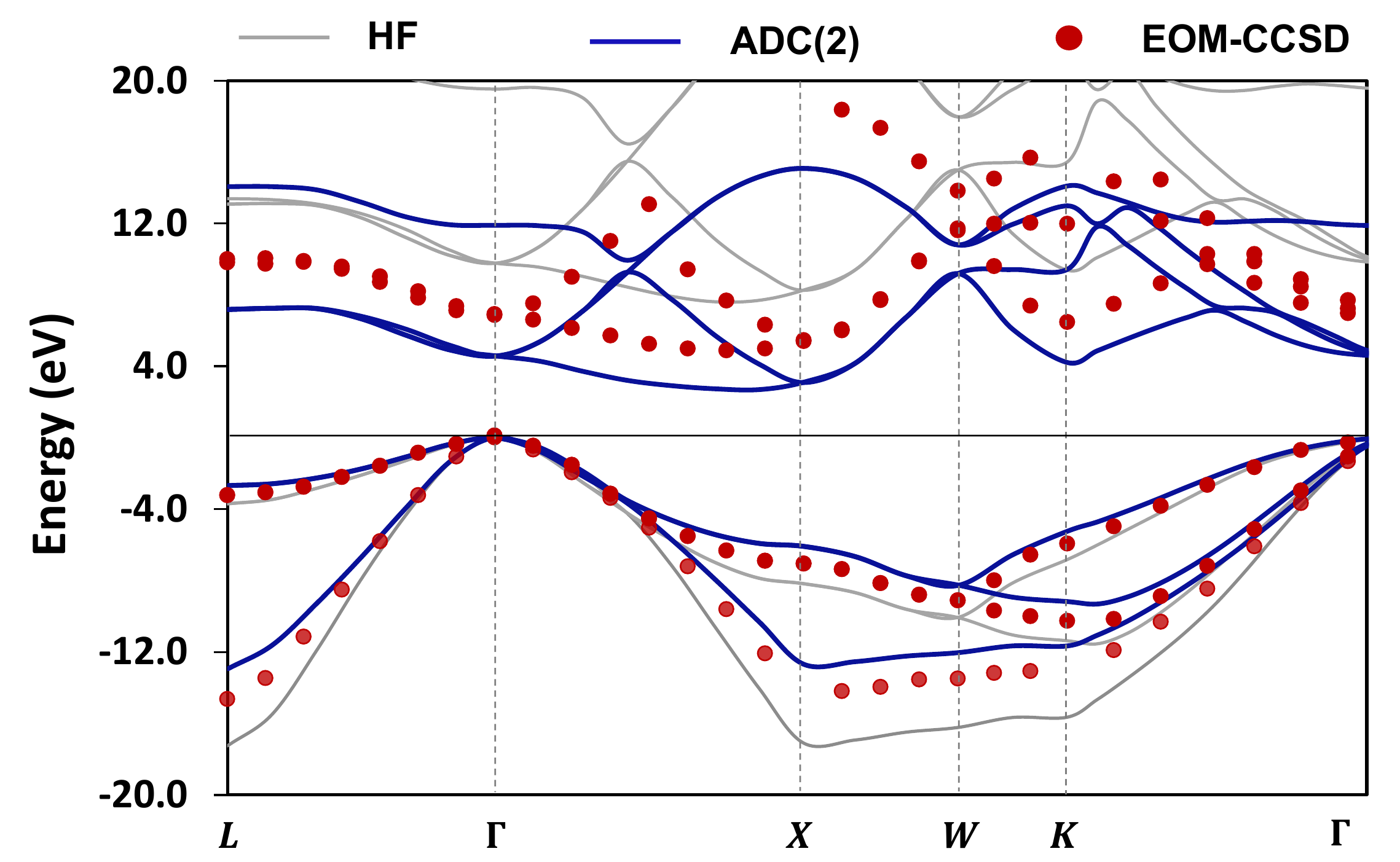}\label{fig:c_eos_2}}  
	\subfigure[]{\includegraphics[width=0.45\textwidth]{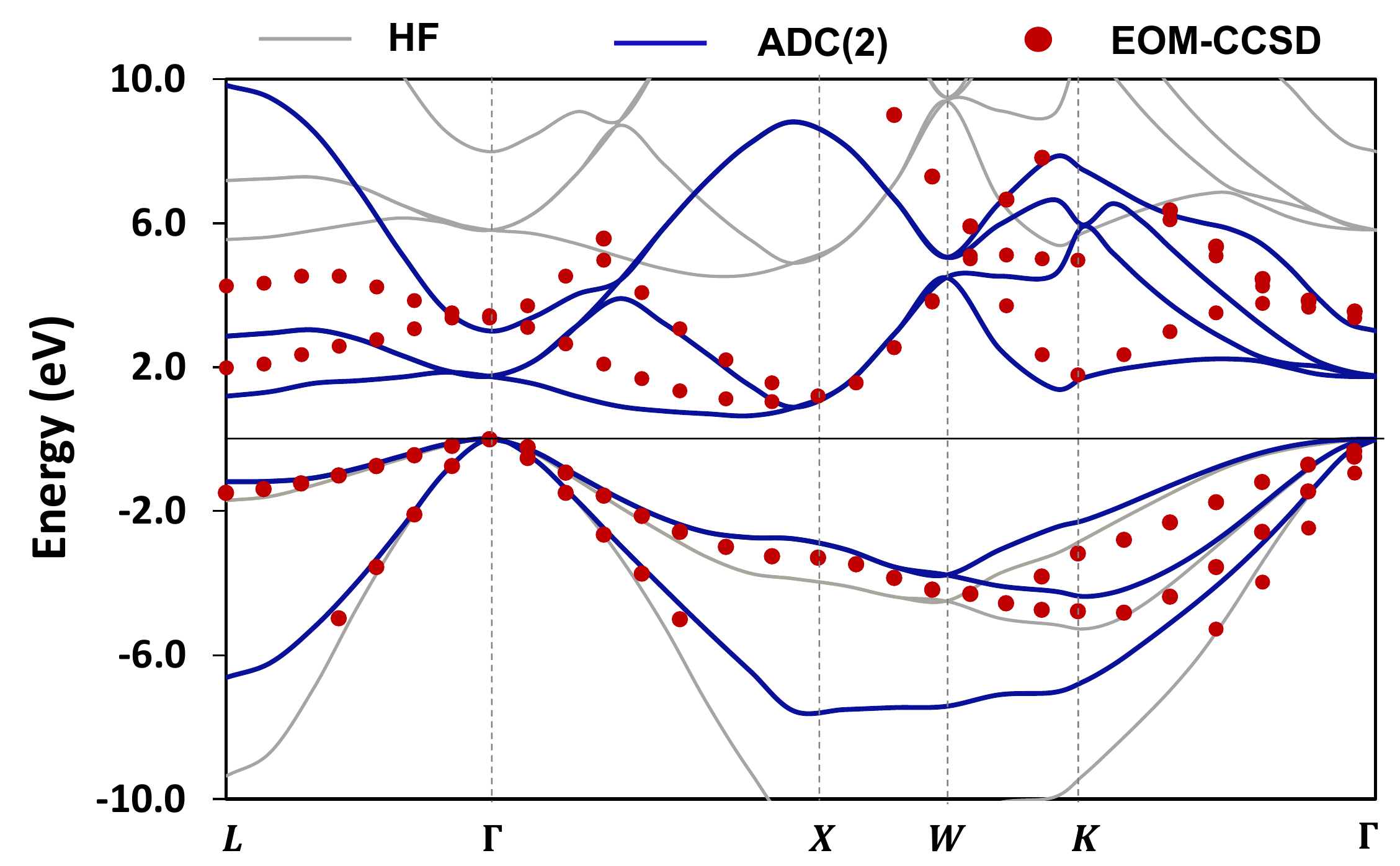}\label{fig:si_eos_2}}  
	\medskip
	\subfigure[]{\includegraphics[width=0.45\textwidth]{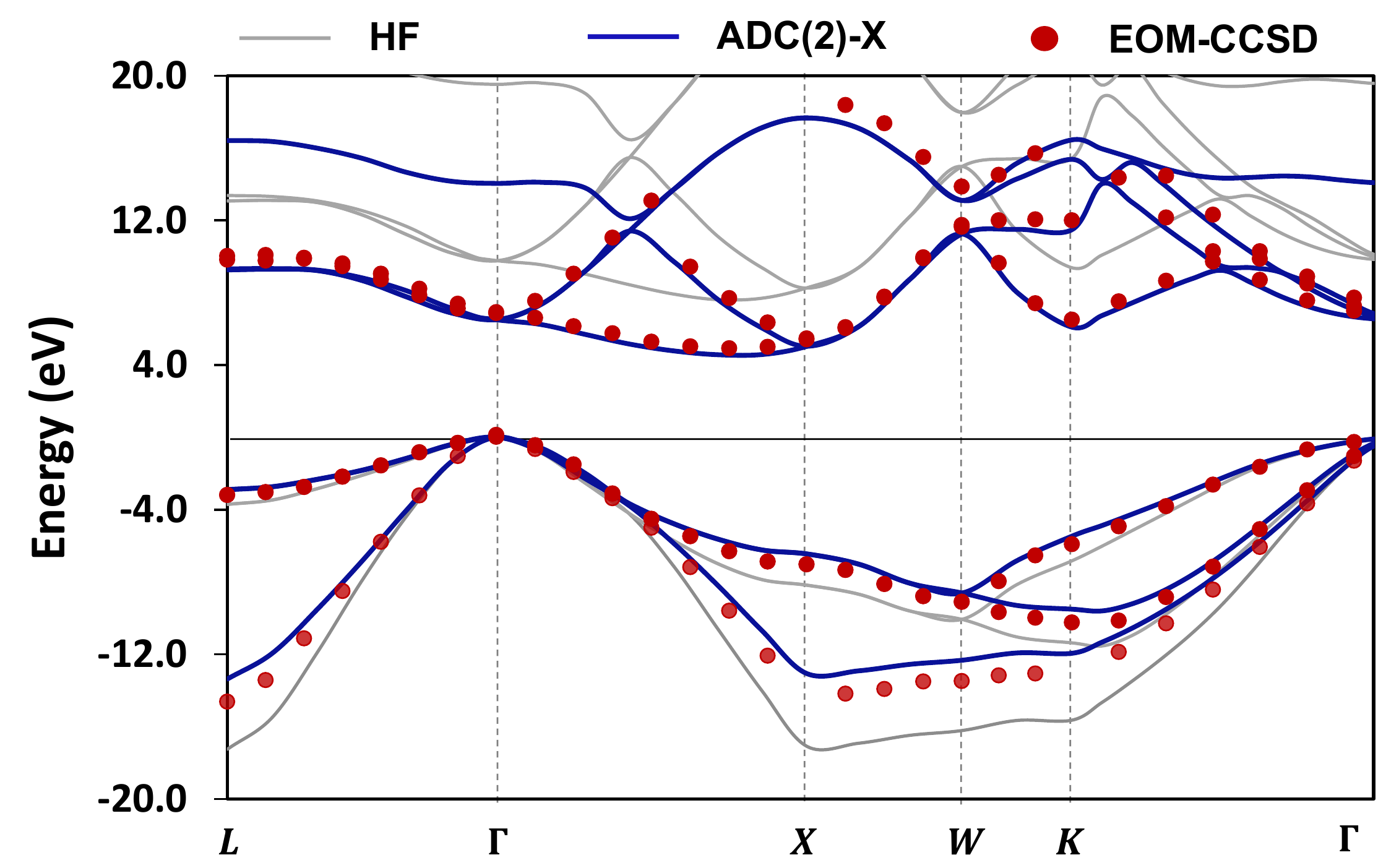}\label{fig:c_eos_2x}}  
	\subfigure[]{\includegraphics[width=0.45\textwidth]{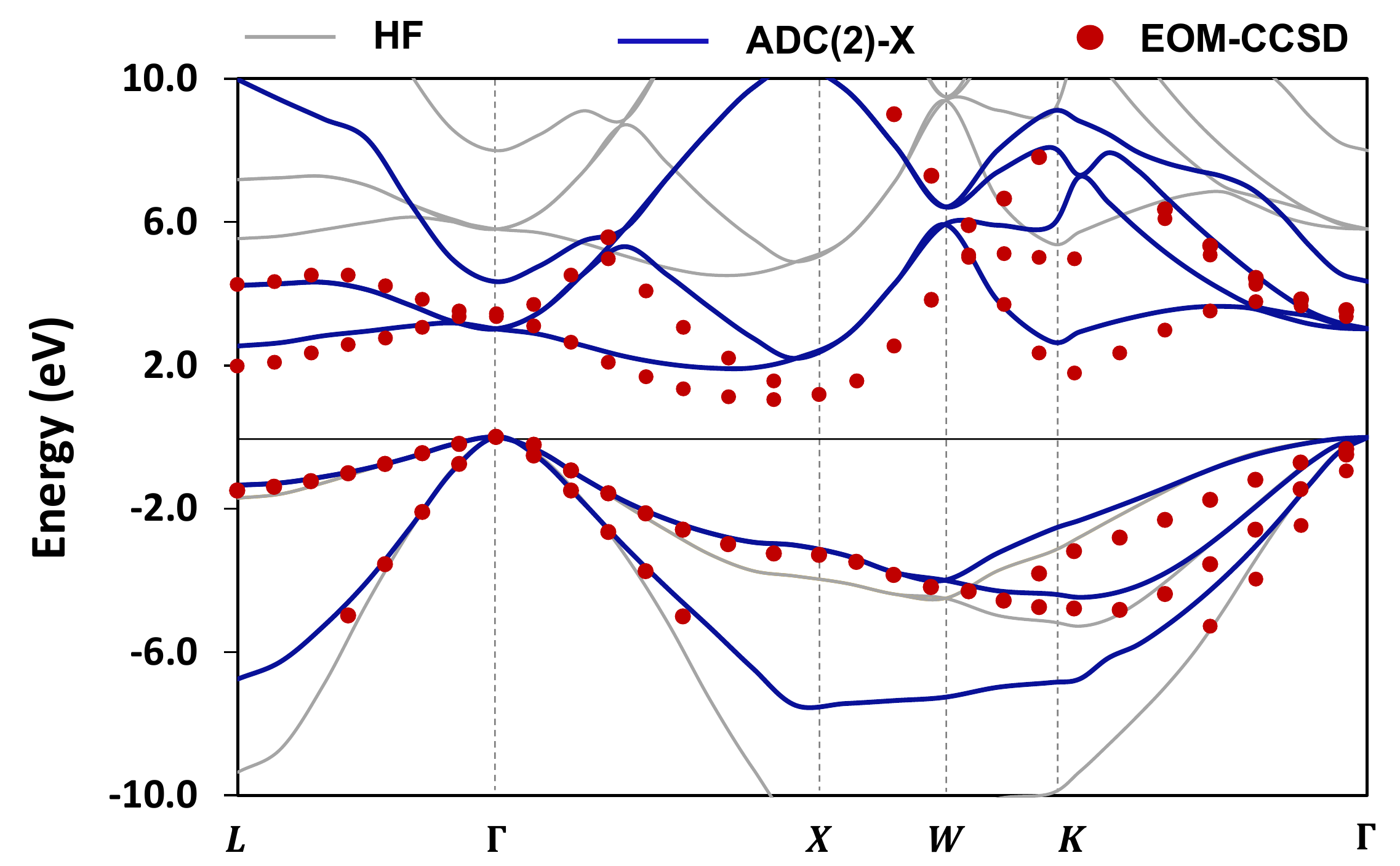}\label{fig:si_eos_2x}}  
	\medskip
	\subfigure[]{\includegraphics[width=0.45\textwidth]{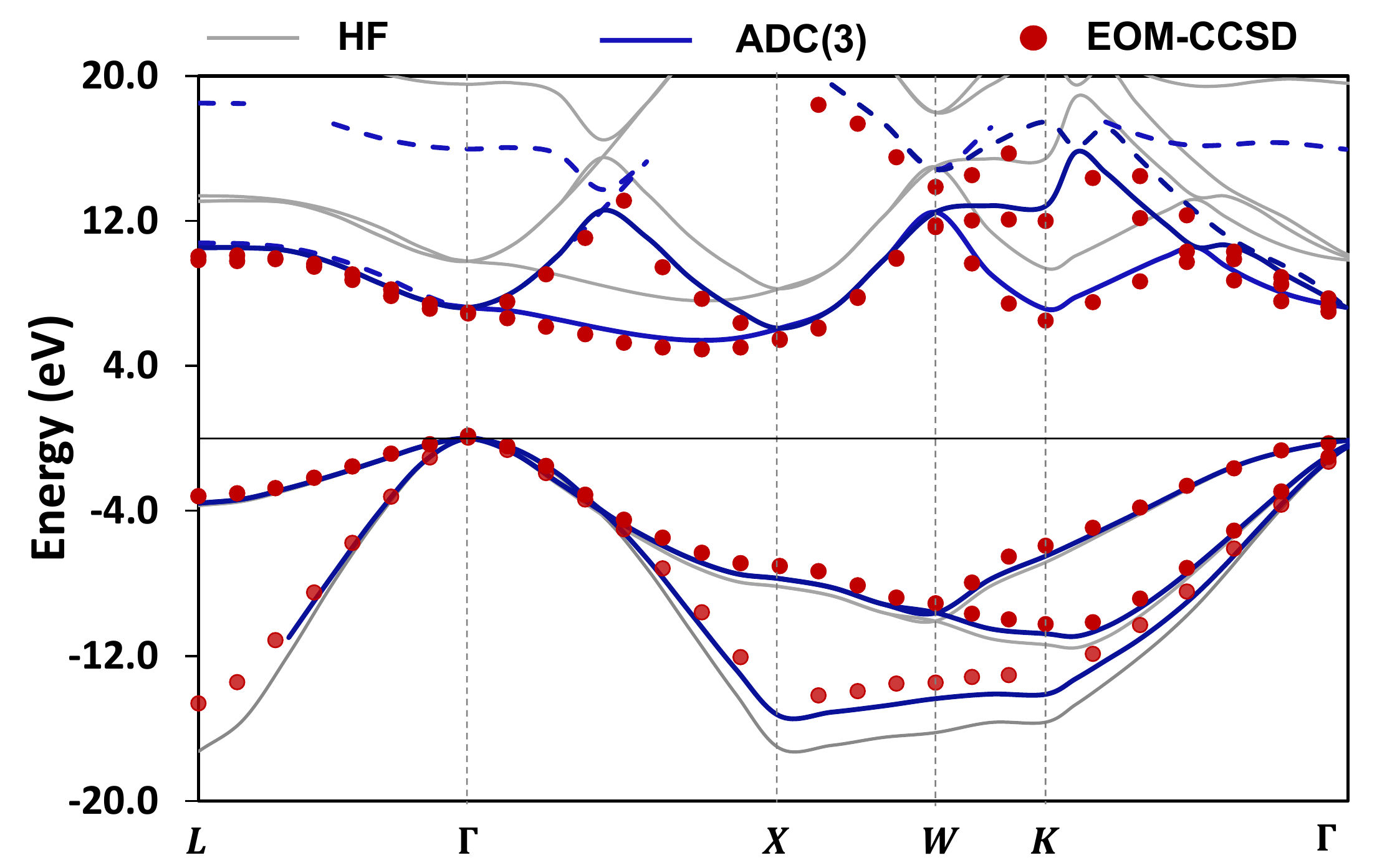}\label{fig:c_eos_3}}  
	\subfigure[]{\includegraphics[width=0.45\textwidth]{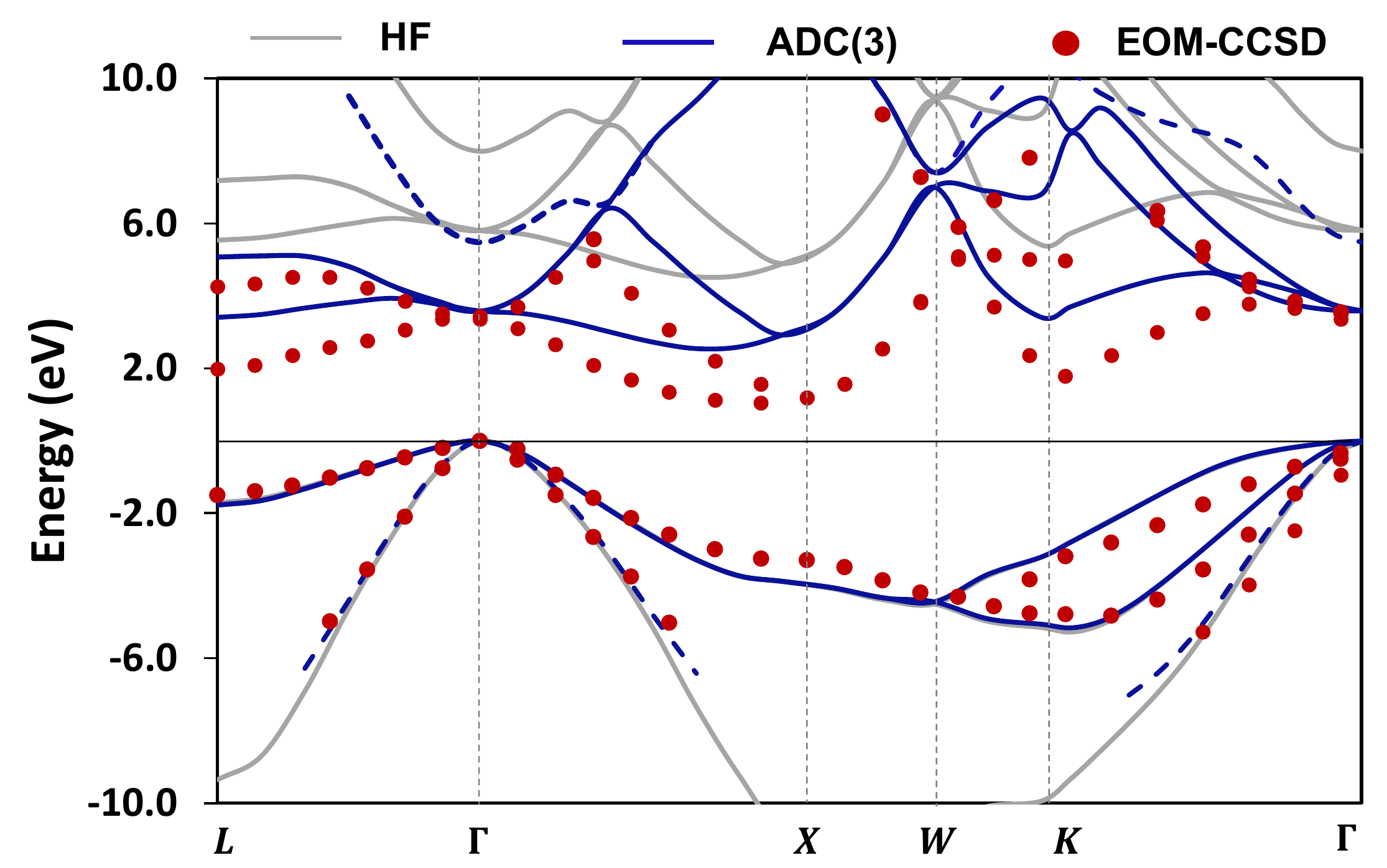}\label{fig:si_eos_3}}
	\caption{Band structures of diamond (a, c, e) and silicon (b, d, f) computed using the periodic SR-ADC methods with $3\times3\times3$ sampling of the Brillouin zone. 
	Results are compared to the band structures calculated using the Hartree--Fock theory (HF) and equation-of-motion coupled cluster theory (EOM-CCSD).\cite{McClain:2017p1209}
	Adapted with permission from Ref.\@ \citenum{Banerjee:2022p5337}. 
	Copyright 2022 American Chemical Society.}
	\label{fig:band_structure}
\end{figure*}

\begin{figure}[t!]
	\centering
	\captionsetup{justification=justified,singlelinecheck=false,font=footnotesize}
	\includegraphics[width=0.4\textwidth]{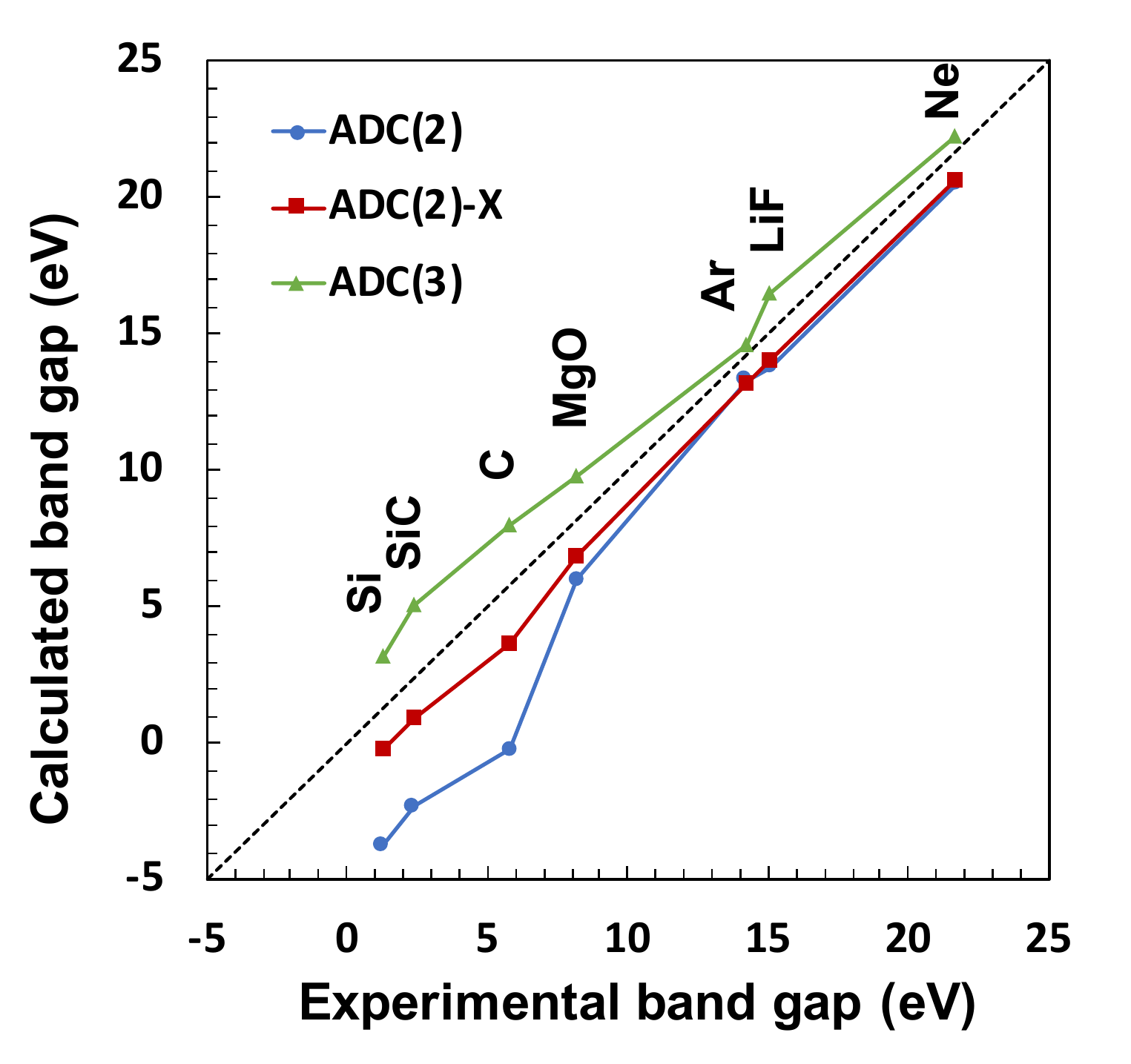}
	\caption{Fundamental band gaps computed using the periodic SR-ADC methods for seven semiconducting and insulating
		materials in comparison to experimental data. 
		The experimental band gaps were corrected to exclude the effects of electron--phonon coupling.
		Reprinted with permission from Ref.\@ \citenum{Banerjee:2022p5337}. 
		Copyright 2022 American Chemical Society.}
	\label{fig:band_gaps}
\end{figure}

When combined with periodic boundary conditions, the ADC methods can be used to elucidate the electronic structure of charged states in chemical systems with translational symmetry.
\cref{fig:band_structure} shows the band structures of solid-state diamond and silicon computed using the periodic implementations of Hartree--Fock, SR-ADC, and EOM-CCSD methods with the $3\times3\times3$ $k$-point sampling of the first Brillouin zone.\cite{Banerjee:2022p5337}
Each band structure plots the energies of charged electronic states as a function of crystal momentum $k$.
For the diamond crystal with experimental band gap of $\sim$ 5.5 eV, SR-ADC(2) correctly reproduces all features of the reference band structure from EOM-CCSD, but underestimates the band gap by $\sim$ 2.3 eV.
The SR-ADC(2)-X and SR-ADC(3) methods improve the description of band gap showing a very good agreement with EOM-CCSD for all points in the band structure.
The silicon crystal with experimental band gap of $\sim$ 1.2 eV proves to be a more challenging test.
In this case, the performance of all SR-ADC methods is non-uniform across different points of the first Brillouin zone, providing evidence that SR-ADC should be used with caution when applied to small-gap semiconductors like silicon.

These results are supported by benchmark data in \cref{fig:band_gaps}, which compares the SR-ADC band gaps extrapolated to thermodynamic limit with the experimental results for seven semiconducting and insulating solids.\cite{Banerjee:2022p5337}
The SR-ADC methods show a good agreement with experiment for large-gap materials such as Ne, LiF, and Ar.
As the experimental band gap decreases, the performance of SR-ADC(2) deteriorates leading to large ($\sim$ 5 to 6 eV) errors for diamond, SiC, and silicon.
SR-ADC(2)-X and SR-ADC(3) significantly improve upon SR-ADC(2) showing errors that range from 1.5 to 2.7 eV relative to experiment.

\subsection{Other properties of charged excited states}
\label{sec:capabilities_accuracy:properties}

The ADC formalism allows to compute a wide range of excited-state properties by providing access to wavefunctions and reduced density matrices of electronically excited states.
Dempwolff and co-workers developed an approach based on intermediate state representation for calculating properties of electron-attached and detached states simulated using SR-ADC. \cite{Dempwolff:2020p024113,Dempwolff:2020p024125,Dempwolff:2021p104117}
An alternative approach based on effective Liouvillian theory has been described by Stahl et al. \cite{Stahl:2022p044106}
To showcase the capabilities of ADC methods for calculating excited-state properties, we highlight the results from these recent studies below.

\subsubsection{Dipole moments}

SR-ADC calculations of dipole moments for electron-attached and ionized states of molecules were reported by Dempwolff et al. \cite{Dempwolff:2020p024113,Dempwolff:2020p024125,Dempwolff:2021p104117}
This work employed three SR-ADC approximations: SR-ADC(2), SR-ADC(3), and SR-ADC(3) with iterative fourth-order treatment of static self-energy (denoted as SR-ADC(3+)).
All three SR-ADC methods predict accurate dipole moments of electron-attached states with mean absolute errors (MAE) of $\sim$ 12 to 16 \% relative to reference data from full configuration interaction.\cite{Dempwolff:2021p104117}
For the ionized states, the computed dipole moments show much stronger dependence on the level of theory.\cite{Dempwolff:2020p024113,Dempwolff:2020p024125}
In this case, the SR-ADC(3+) method yields the most reliable results (MAE = 21 \%), while the average error increases two-fold for SR-ADC(2)  and SR-ADC(3).
We note that in these studies the calculations of dipole moments were performed using the one-particle reduced density matrices evaluated up to the second order in single-reference perturbation theory.
The importance of third-order correlation effects in the calculations of SR-ADC(3) and SR-ADC(3+) dipole moments remains to be studied.

\subsubsection{Dyson orbitals and electronic density differences}

\begin{figure}[t!]
	\captionsetup{justification=justified,singlelinecheck=false,font=footnotesize}
	\subfigure[]{\includegraphics[width=0.4\textwidth]{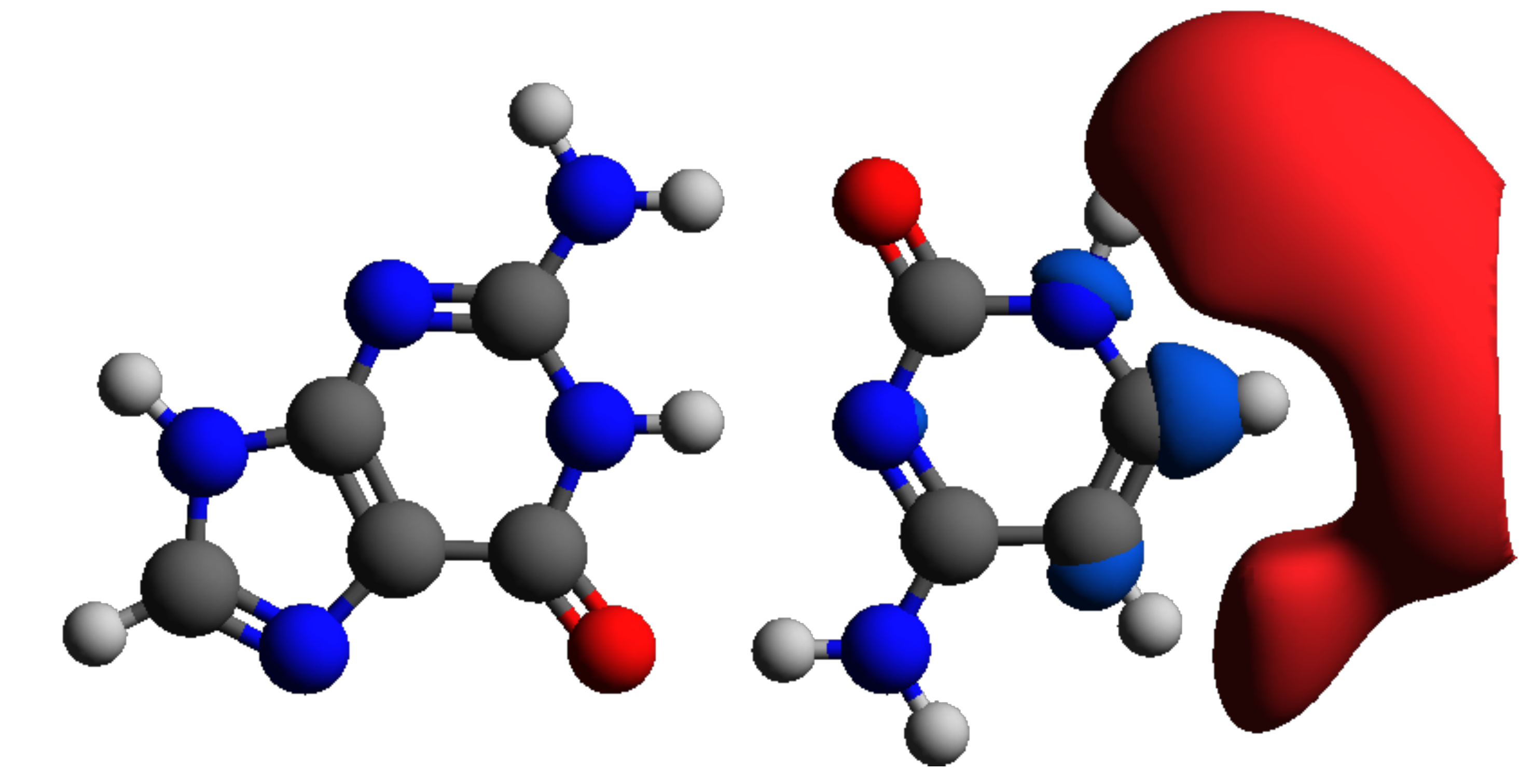}\label{fig:gc_dyson_neutral}} \qquad
	\subfigure[]{\includegraphics[width=0.4\textwidth]{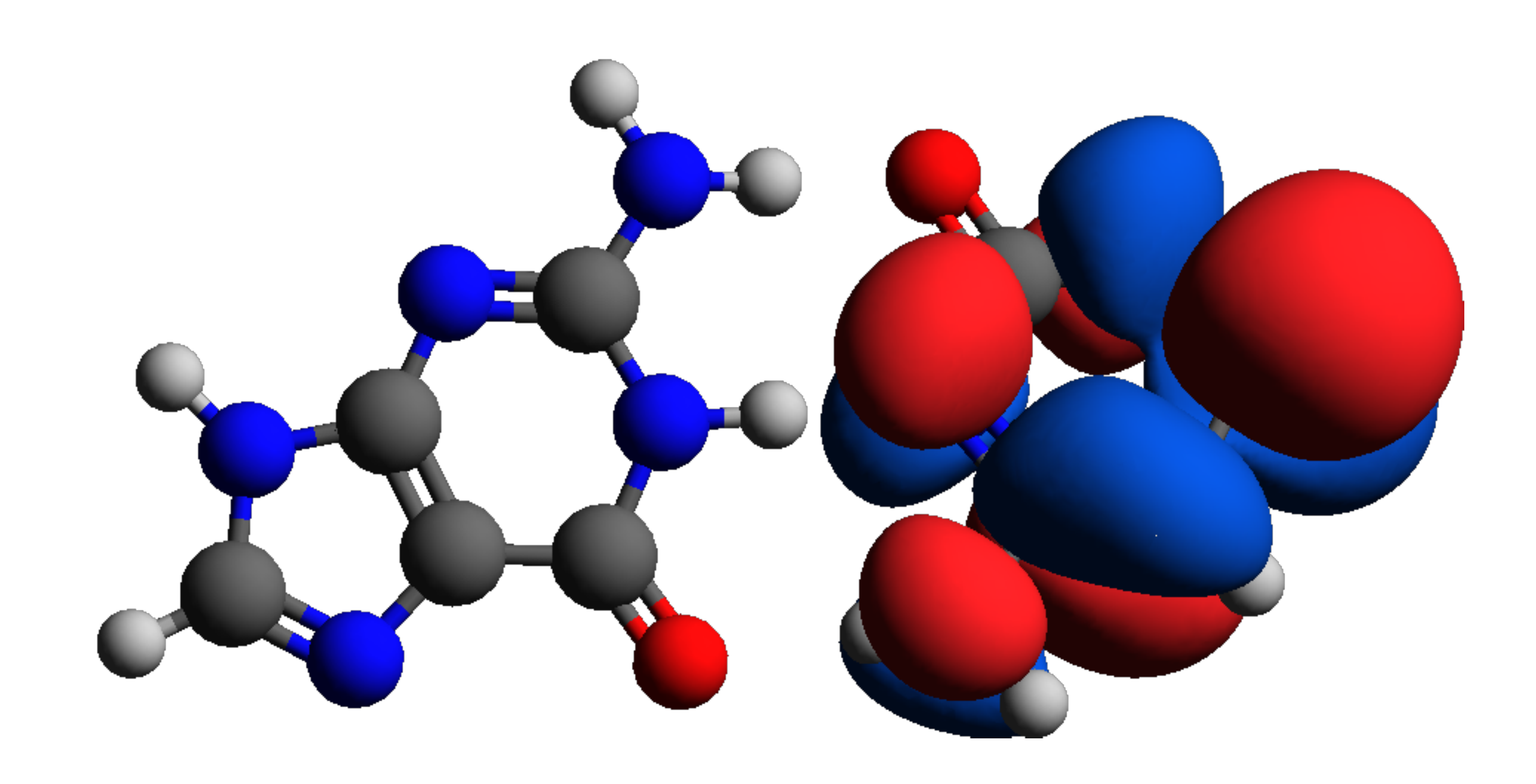}\label{fig:gc_dyson_anion}} 
	\caption{Dyson orbitals for the lowest-energy electron-attached state of the guanine--cytosine base pair at the neutral (a) and anion (b) equilibrium geometries computed using the SR-ADC(3) method. 
	Reprinted from Ref.\@ \citenum{Banerjee:2021p074105}, with the permission of AIP Publishing.
	}
	\label{fig:gc_dyson}
\end{figure}

\begin{figure*}[t!]
	\captionsetup{justification=justified,singlelinecheck=false,font=footnotesize}
	\subfigure[]{\includegraphics[width=0.4\textwidth]{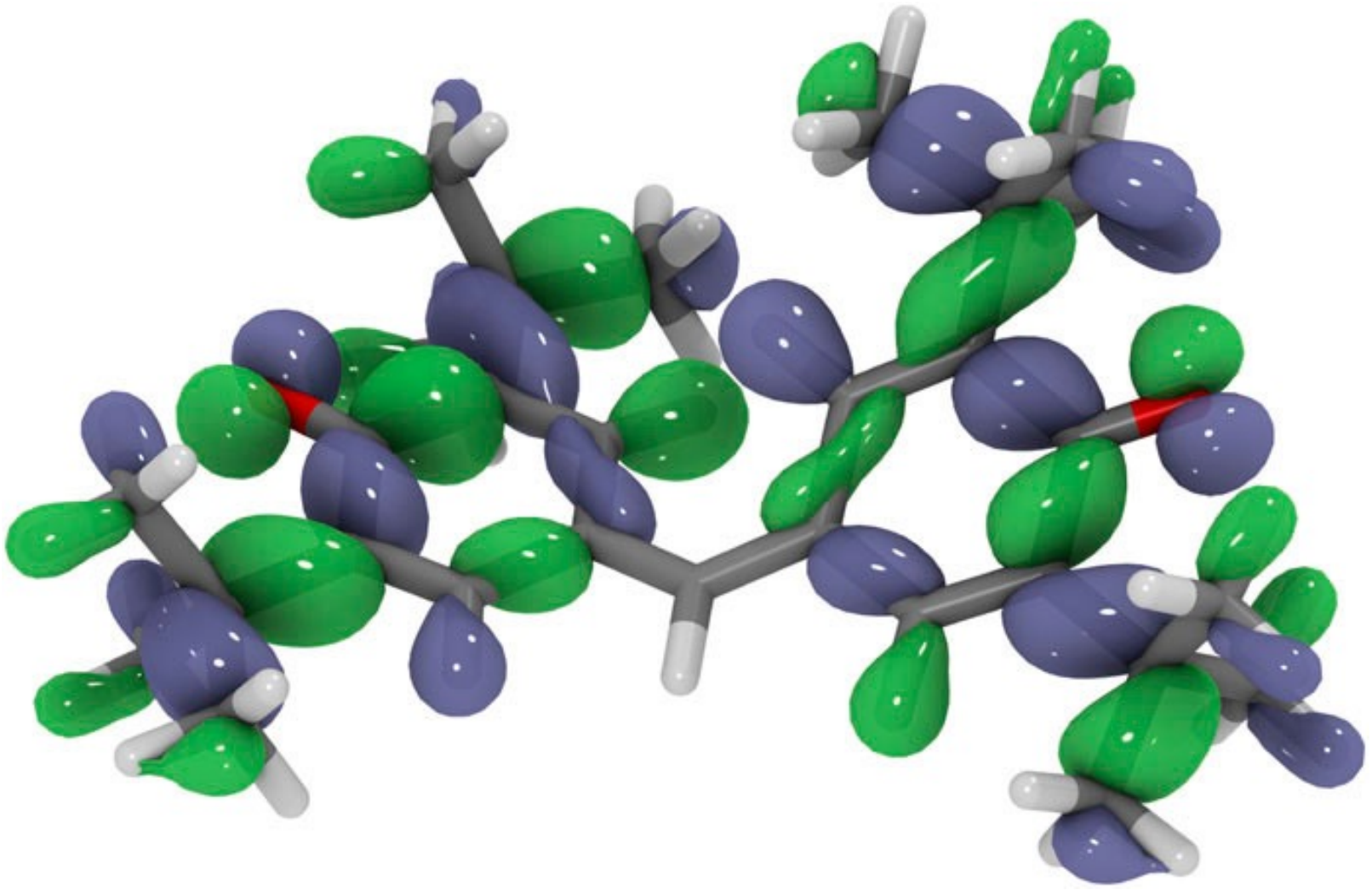}\label{fig:dyson_orb_gfr}} \qquad
	\subfigure[]{\includegraphics[width=0.4\textwidth]{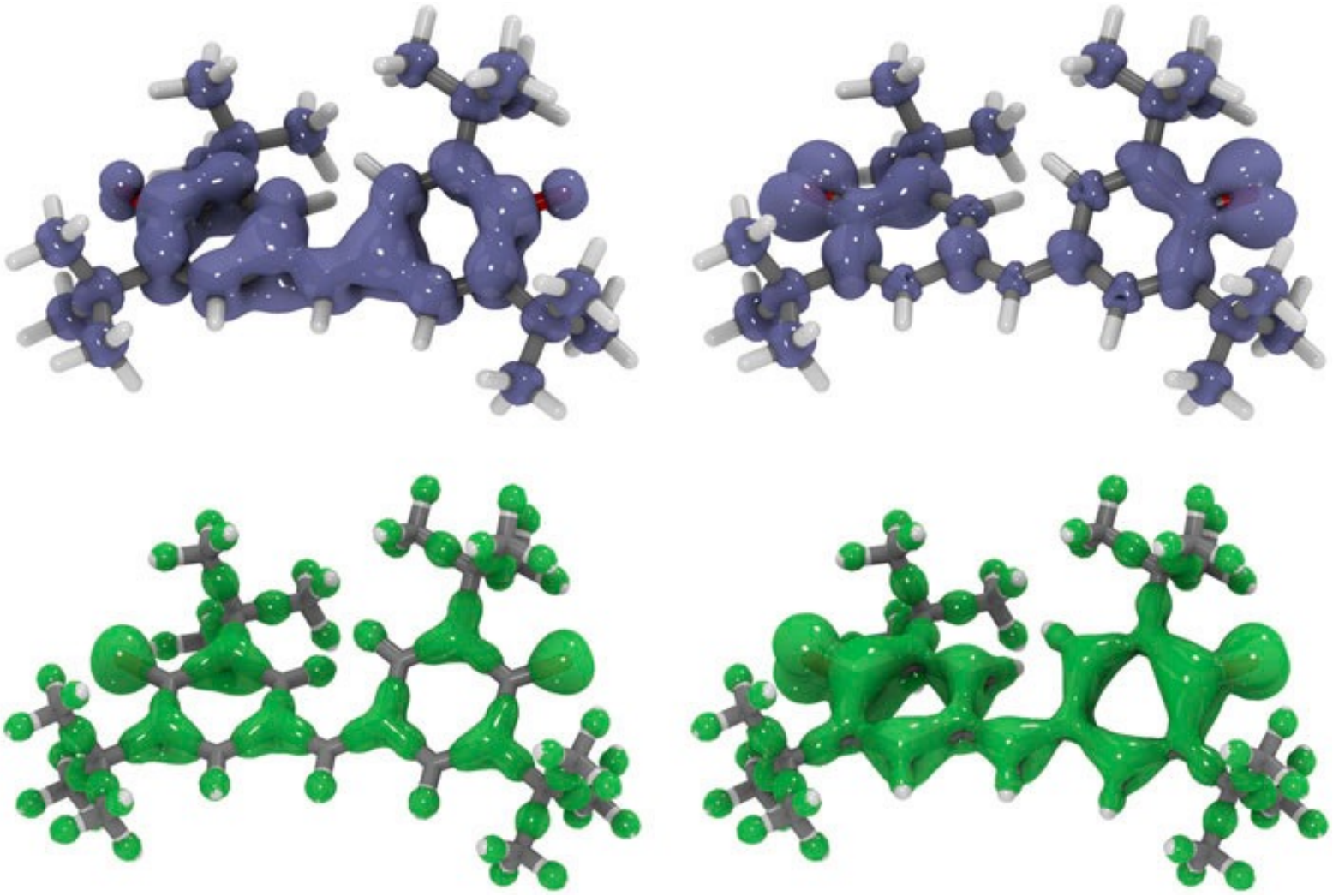}\label{fig:den_dif_gfr}} 
	\caption{(a) Dyson orbital for a satellite transition in the photoelectron spectrum of galvinoxyl free radical (GFR) cation computed using SR-ADC(3).
	(b) The same transition is represented using detachment (upper row) and attachment (lower row) densities, with the $\alpha$ (spin up) contributions shown on the left, and the $\beta$ (spin down) contributions shown on the right. 
	Reprinted from Ref.\@ \citenum{Dempwolff:2020p024113}, with the permission of AIP Publishing.
	}
	\label{fig:den_diff}
\end{figure*}

The ADC methods enable direct calculations of Dyson orbitals (\cref{eq:Dyson_orb}) that provide information about the spatial localization of an electron or hole created as a result of charged excitation and can be helpful in interpreting transitions in photoelectron spectra.\cite{Melania:2007p234106,Ortiz:2020p070902} 
\cref{fig:gc_dyson} shows Dyson orbitals for electron attachment to guanine--cytosine DNA base pair computed at the neutral ground-state (a) and anion (b) equilibrium geometries using SR-ADC(3). \cite{Banerjee:2021p074105}
At the neutral state geometry, Dyson orbital reveals the dipole-bound nature of electron-attached state with excess electron localized outside the molecule. 
Allowing the anion geometry to relax leads to a valence-bound state with electron localized on the $\pi$-orbitals of cytosine molecule. 

As one-electron functions, Dyson orbitals are not able to describe charged excitations involving two or more electrons, such as low-intensity satellite transitions in photoelectron spectra, which are often difficult to interpret without the help from theoretical calculations. 
In this case, a more accurate picture of a charged excitation is provided by visualizing the change in electronic density between the ground and charged-excited state, which can be calculated from the ADC excited-state reduced density matrices and decomposed into detachment and attachment density contributions.\cite{Plasser:2014p024106}
This is illustrated in \cref{fig:den_diff}, which compares the SR-ADC(3) Dyson orbital with detachment and attachment densities for a satellite transition of galvinoxyl free radical (GFR).\cite{Dempwolff:2020p024113}
Although the Dyson orbital suggests that this transition should be interpreted as removing an electron from the $\sigma$-orbitals of GFR, the density difference reveals a more complicated process with an $n \rightarrow \pi^{*}$ rearrangement of $\beta$-electron density and a $ \pi \rightarrow \sigma^{*}$ excitation of $\alpha$-electrons resulting in a positively charged hole with predominantly $\pi$-character.

\subsubsection{Spin properties}

\begin{figure*}[t!]
	\captionsetup{justification=justified,singlelinecheck=false,font=footnotesize}	
	\subfigure[]{\includegraphics[width=0.7\textwidth]{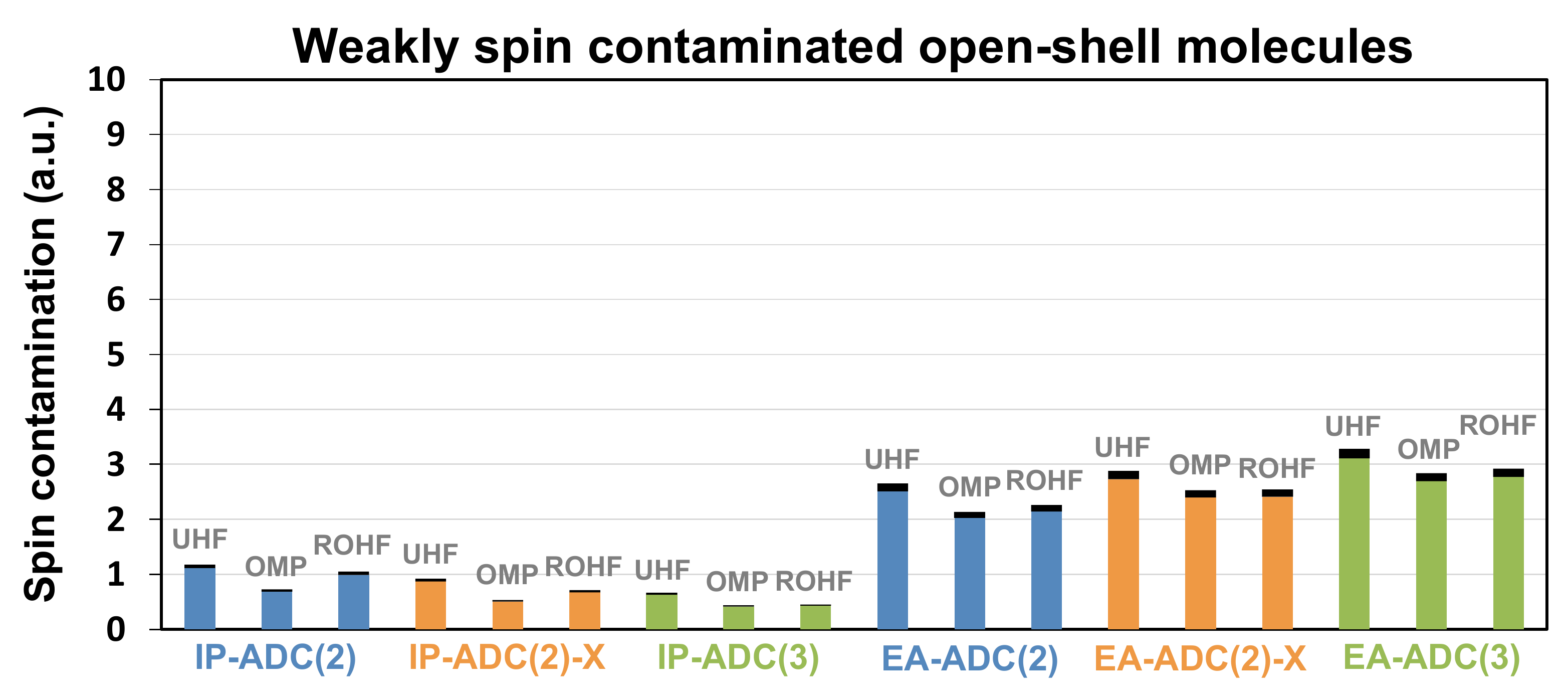}\label{fig:wc_sc}}  
	\subfigure[]{\includegraphics[width=0.7\textwidth]{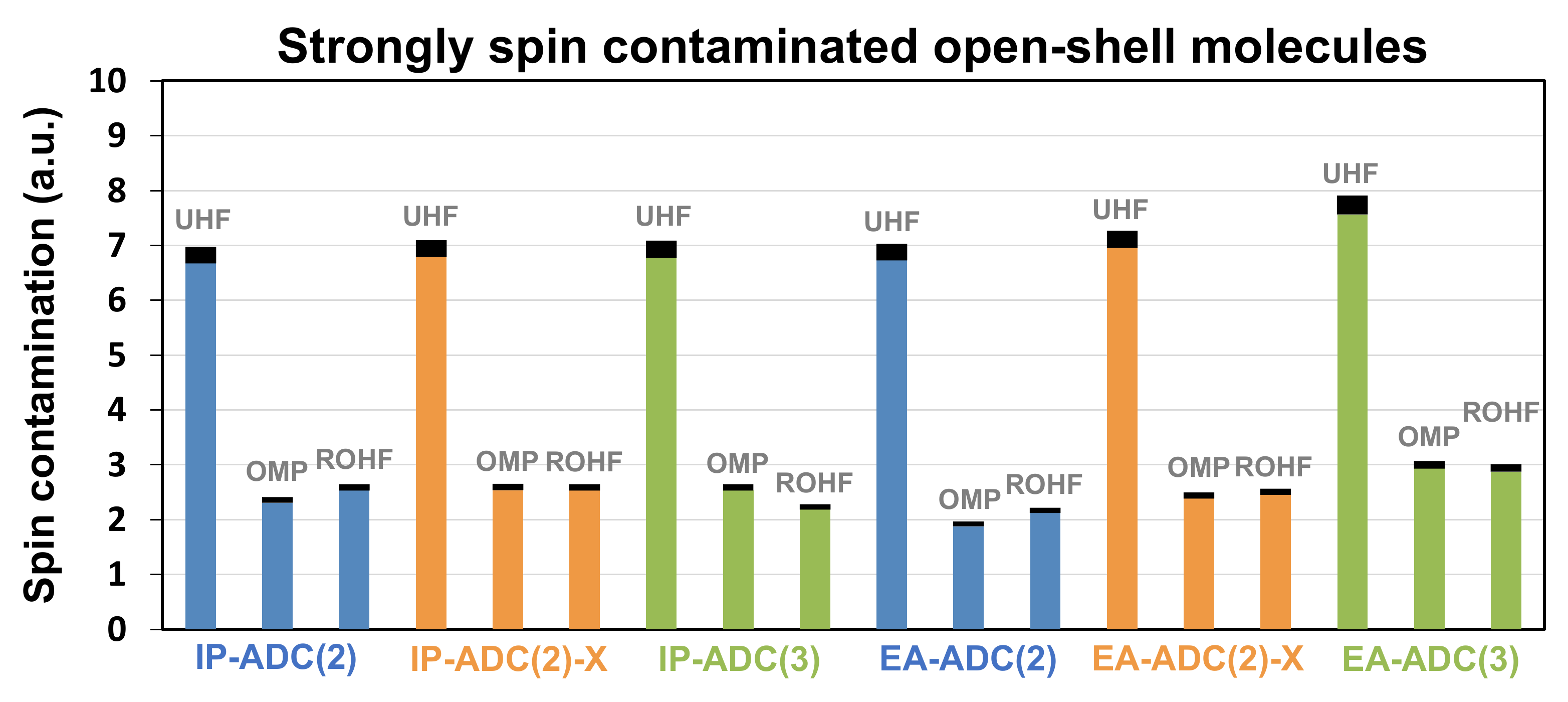}\label{fig:sc_sc}}  
	\caption{Spin contamination in the lowest-energy ionized and electron-attached states for (a) 18 weakly and (b) 22 strongly spin-contaminated molecules computed using SR-ADC with three different reference wavefunctions. 
		Colored boxes indicate the sum of spin contamination for all molecules in each set, black boxes show the average spin contamination.
		Adapted from Ref.\@ \citenum{Stahl:2022p044106}, with the permission of AIP Publishing.}
	\label{fig:spin}
\end{figure*}

The ADC reduced density matrices can be used to compute excited-state expectation values for a variety of important observable properties.
Recently, Stahl et al.\@ investigated the accuracy of SR-ADC methods for simulating spin properties in charged states of open-shell molecules. \cite{Stahl:2022p044106}
They demonstrated that SR-ADC(2), SR-ADC(2)-X, and SR-ADC(3) exhibit non-negligible spin contamination in the calculated excited states.
\cref{fig:spin} shows that the excited-state spin contamination is relatively insensitive to the order of SR-ADC approximation and is particularly large for molecules with strong spin contamination in the UHF reference wavefunction ($\Delta S^2$ $>$ 0.1 a.u.). 
Performing SR-ADC calculations with the restricted open-shell Hartree--Fock or orbital-optimized M\o ller--Plesset reference wavefunctions significantly reduces the excited-state spin contamination, although does not cure it entirely.
As shown in \cref{fig:sc}, these improvements in describing spin properties lead to a noticeable improvement of the SR-ADC(3) accuracy for charged excitation energies, while the performance of SR-ADC(2) and SR-ADC(2)-X is affected to a much lesser extent (see \cref{sec:capabilities_accuracy:energies} for more information).\cite{Stahl:2022p044106}

\section{Summary and Outlook}
\label{sec:summary_outlook}

In this perspective, we reviewed the current state of algebraic diagrammatic construction (ADC) theory for simulating charged excited states, focusing primarily on the non-Dyson formulation of this theoretical approach.
ADC offers a compromise between accuracy of its approximations and computational cost, providing a hierarchy of systematically improvable and size-consistent methods for simulating many excited electronic states at the same time. 
The ADC methods allow to compute spectroscopic observables (excitation energies and intensities in photoelectron spectra) and excited-state properties (e.g., electronic density differences, dipole moments, spin) that can help assigning features in experimental spectra, obtaining deeper insight into excited-state electronic structure, and making reliable predictions of spectroscopic properties for future experiments. 
Recent developments of periodic and multireference variants of ADC have enabled applications of this theoretical approach to crystalline solids and chemical systems with complex, multiconfigurational electronic structures.

Looking ahead, the capabilities of ADC theory for simulating charged excited states can be expanded in many different directions. 
Applications to large molecular or crystalline systems require lowering the computational cost of ADC methods, which can be achieved by using local correlation,\cite{Schutz:2001p661,Werner:2003p8149,Riplinger:2013p034106} frozen natural orbital,\cite{Taube:2005p837,Mester:2018p094111} or tensor factorization techniques.\cite{Hohenstein:2012p044103,Parrish:2012p224106,Hohenstein:2012p221101} 
Simulating charged excitations in realistic reaction environments necessitate incorporating environmental and solvation effects.\cite{Scheurer:2018p4870} 
Characterizing potential energy surfaces of charged states would benefit from the implementation of analytical gradients. 
Additional developments are also needed for simulating core-ionized states and X-ray photoelectron spectra where incorporating spin--orbit coupling effects can be important.
Lastly, improved periodic ADC methods are necessary for accurate calculations of charged excitations in small-gap semiconductors where multireference effects play significant role.
Expanding these horizons will create new opportunities to advance our understanding of charged electronic states using ADC theory.

\acknowledgement
	Acknowledgment is made to the donors of the American Chemical Society Petroleum Research Fund for support (or partial support) of this research. (PRF \#65903-ND6). 
	This work was also supported by the National Science Foundation, under Grant No. CHE-2044648.
	Additionally, S.B. was supported by a fellowship from Molecular Sciences Software Institute under NSF Grant No.\@ ACI-1547580.
	Many calculations reported here were performed at the Ohio Supercomputer Center under projects PAS1583 and PAS1963.\cite{OhioSupercomputerCenter1987} 


\providecommand{\latin}[1]{#1}
\makeatletter
\providecommand{\doi}
  {\begingroup\let\do\@makeother\dospecials
  \catcode`\{=1 \catcode`\}=2 \doi@aux}
\providecommand{\doi@aux}[1]{\endgroup\texttt{#1}}
\makeatother
\providecommand*\mcitethebibliography{\thebibliography}
\csname @ifundefined\endcsname{endmcitethebibliography}
  {\let\endmcitethebibliography\endthebibliography}{}

\end{document}